\documentclass[10pt]{iopart}
\usepackage{graphicx}
\usepackage{iopams}  

\newcommand{\be}{\begin{equation}}
\newcommand{\ba}{\begin{eqnarray}}
\newcommand{\ee}{\end{equation}}
\newcommand{\ea}{\end{eqnarray}}
\newcommand{\ignore}[1]{}

\newcommand{\q}{\bi{q}}
\newcommand{\APL}{Appl.\ Phys.\ Lett.}

\begin{document}

\topical{Disorder effects in diluted magnetic semiconductors}

\author{Carsten Timm}

\address{Institut f\"ur Theoretische Physik, Freie Universit\"at
Berlin, Arnimallee 14, D-14195 Berlin, Germany}

\begin{abstract}
In recent years, disorder has been shown to be crucial for the
understanding of diluted magnetic semiconductors. Effects of disorder in
these materials are reviewed with the emphasis on theoretical works. The
types and spatial distribution of defects are discussed. The effect of
disorder on the intimately related transport and magnetic properties are
considered from the viewpoint of both the band picture and the
isolated-impurity approach. Finally, the derivation and properties of
spin-only models are reviewed.
\end{abstract}



\ead{timm@physik.fu-berlin.de}


\section{Introduction}
\label{sec.intro}

Diluted magnetic semiconductors (DMS) are promising materials for
applications as well as interesting from the basic-physics point of view.
Possible applications exist in spin electronics (\emph{spintronics})
\cite{WAB01,D.why}, which employ the spin degree of freedom of electrons in
addition to their charge. This may allow the incorporation of ferromagnetic
elements into semiconductor devices, and thus the integration of data
processing and magnetic storage on a single chip. Since the electronic spin
is a quantum-mechanical degree of freedom, quantum-interference effects
could be exploited in devices, eventually leading to the design of quantum
computers \cite{QC}.

DMS also pose a number of questions of fundamental interest for
condensed-matter physics: What is the nature of disorder? What role does it
play for transport and magnetism? What is the interplay between disorder
physics and strong correlations? What is the mechanism of ferromagnetic
ordering? What is the interplay between transport and magnetic properties?
The present topical review concentrates on the first two questions. A lot
of work has been done on disorder effects in nonmagnetic semiconductors and
metals \cite{LeR85,KrM93,Imr97}. Only during the last few years disorder
effects in DMS have been considered. They can be expected to be strong due
to the presence of a high concentration of charged impurities; the typical
distance between these defects is roughly of the same order as the Fermi
wave length. Possibly Wan and Bhatt \cite{WaB00} were the first to
emphasize the importance of disorder for DMS. We here apply the term
``disorder effects'' to those effects that really depend on the
\emph{random spatial distribution} of impurities and not only on their
\emph{presence}, such as the change of the carrier concentration.

This paper attempts to give an overview over present theoretical approaches
for disorder in DMS. The emphasis is on III-V materials such as GaAs and
InAs doped with manganese since these have been studied most intensively
and also show promisingly high Curie temperatures. There are a number of
excellent review articles covering aspects of the physics of DMS not
discussed here. Discussions of the motivation for studying DMS drawn from
possible applications can be found in \cite{WAB01} and \cite{D.why}.
Ferromagnetism in III-V materials is discussed by Ohno \cite{Ohn98,Ohn99}
and Ohno and Matsukura \cite{OhM01} emphasizing experimental properties and
by Dietl and Ohno \cite{DiO01} discussing both experimental and theoretical
aspects. The theoretical results are based on the Zener model and a
mean-field approximation for the magnetic order. K\"onig \etal \cite{KSJ03}
provide a more extensive review of theoretical results obtained by this
approach. Lee \etal \cite{LJM02} discuss DMS heterostructures for possible
applications. It may be useful to mention the ``Ferromagnetic Semiconductor
Spintronics Web Project'' at http://unix12.fzu.cz/ms/index.php. This web
page contains a large body of theoretical and experimental results and an
extended bibliography. Sanvito \etal \cite{STH02} review \textit{ab-initio}
calculations for III-V DMS. Dietl \cite{Die02} and Pearton \etal
\cite{PAO03} give overviews over ferromagnetic semiconductors not limited
to diluted III-V compounds.

The remainder of this paper is organized as follows: In the rest of 
\sref{sec.intro} we first discuss basic properties of DMS, then the types
of impurities thought to be relevant and finally general consequences of
the presence of magnetic impurities. In \sref{sec.prop} we consider the
disorder more carefully, in particular the spatial distribution of
impurities. \Sref{sec.trans} concerns the effect of disorder on transport
properties and \sref{sec.mag} deals with disorder effects on magnetism.
Finally, \sref{sec.conc} draws some conclusions and lists open questions.

\subsection{Basic properties of DMS}

Diluted magnetic semiconductors are realized by doping a semiconducting
host material with magnetic ions, typically manganese. The basic properties
of DMS depend on the type of host material: For III-V semiconductors such
as GaAs, manganese plays a dual role in that it acts as an acceptor and
provides a localized spin due to its partially filled (in this case
half-filled) d-shell. Manganese in GaAs is in a $\mathrm{Mn}^{2+}$ state
\cite{LJM97}. The doped hole is weakly bound to the acceptor by Coulomb
attraction, forming a shallow impurity state split off from the
valence-band top \cite{LJM97,OKR98}. The local impurity spin is coupled to
the valence-band holes by exchange interaction. The shallow-impurity
description probably applies since states dominated by manganese d-orbitals
are far from the Fermi energy \cite{LJM97,OKR98,BKD02,Die02}. However, see
\cite{SMR02,KSH02} for conflicting views. \textit{Ab-initio} theory does
not yet consistently support the shallow-acceptor picture, see
\sref{sus.types}. The situation for manganese in InAs, GaSb, InSb and AlSb
is similar \cite{DMO02,Die02}. On the other hand, in GaN manganese forms a
\emph{deep} acceptor probably due to significant admixture of d-orbitals
\cite{Zunger,DMO02,MaZ03a}.

Since manganese acts as an acceptor, manganese-doped III-V semiconductors
are of p-type. However, the observed hole concentration is lower than the
concentration of manganese impurities due to compensation, probably by
arsenic antisites (arsenic ions substituted for gallium) and manganese
interstitials. Both types of defects provide electrons, \textit{i.e.}, they
are double donors.

The experimental determination of the hole concentration is not trivial
\cite{RSB03}. It is usually obtained from the ordinary Hall coefficient
\cite{MOS98,Ohn98,Ohn99,RSB03}, measured in a strong external field to
saturate the \emph{anomalous} Hall effect, which is due to the
magnetization. But even for large fields the Hall resistivity is not linear
in field. Careful analysis nevertheless finds agreement with the carrier
concentration obtained independently from electrochemical profiling
\cite{RSB03}. The Raman scattering intensity has also been used
successfully to obtain the carrier density \cite{SCC02}.

Since the equilibrium solubility of manganese is very low, III-V DMS have
to be grown under non-equilibrium conditions. Mostly, molecular beam
epitaxy (MBE) at low temperatures of the order of $250^\circ\mathrm{C}$ has
been used but metalorganic vapour phase epitaxy (MOVPE) at higher
temperatures has also been employed. Ferromagnetism in III-V DMS has been
first found in (In,Mn)As grown by MBE \cite{OMP92}, but the Curie
temperature is low. In (Ga,Mn)As ferromagnetism above $100\:\mathrm{K}$ has
first been demonstrated by Ohno \etal \cite{OSM96}. The total
magnetization is dominated by the impurities because of compensation and
since the impurity spins are larger ($S=5/2$ in this case). Recently Curie
temperatures around $160\:\mathrm{K}$ have been reached partly due to
improved control over the concentration of defects
\cite{Schiffer,Gallagher}. In (In,Mn)As grown by MOVPE a higher Curie
temperature of $333\:\mathrm{K}$ has been reported \cite{BlW02,BlW03}. This
result is not well understood since the Curie temperature is found to be
independent of manganese concentration. Also, the magnetization has been
measured in an applied field much larger than the coercive field
\cite{BlW02,BlW03} so that fluctuations are strongly suppressed and the
measured $T_{\mathrm{c}}$ is not the true transition temperature where
spontaneous long-range order appears. In (Ga,Mn)N even higher Curie
temperatures in excess of $750\:\mathrm{K}$ have been observed
\cite{SSS02}. However, the failure to observe an anomalous Hall effect
indicates that ferromagnetism and electronic conduction might take place in
different phases.

In II-VI materials manganese is isovalent with the host cation so that it
only provides a spin. The interplay of Coulomb attraction and exchange
important for manganese in III-V compounds is thus missing here. Additional
doping is required to introduce charge carriers. Manganese-doped II-VI
semiconductors have been studied quite extensively
\cite{Furdynarev,Dietl.26}, mostly before the advent of III-V materials
with high $T_{\mathrm{c}}$. II-VI DMS are typically not ferromagnetic but
ferromagnetism has been achieved using modulation doping \cite{HWA97}. The
Curie temperature does not exceed $2\:\mathrm{K}$.

Regarding group-IV semiconductors, only Ge has been successfully doped with
magnetic ions, in this case manganese, resulting in Curie temperatures of
up to $116\:\mathrm{K}$ \cite{GeMn}. This DMS is strongly insulating
\cite{GeMn}. It is similar to III-V compounds in that manganese acts
as a double acceptor, leading to a strongly p-type semiconductor.

Magnetically doped wide-gap oxide semiconductors have been predicted to
show high Curie temperatures \cite{DOM00}. Matsumoto \etal
\cite{MMS01} have prepared (Ti,Co)O$_2$ with cobalt concentrations of up to
8\%. Curie temperatures above $400\:\mathrm{K}$ have been estimated from the
experiments \cite{MMS01}. Recent measurements of the anomalous Hall effect
show that the carriers are strongly affected by the magnetic order
\cite{TFY03}. However, oxide DMS are still not well understood.
A brief review can be found in \cite{FYT03}.

For a DMS to order ferromagnetically there has to be an effective
ferromagnetic interaction between the impurity spins, which is very
probably carrier-mediated for most or all ferromagnetic DMS. This picture
is supported by mea\-sure\-ments of the anomalous Hall effect in (Ga,Mn)As
\cite{MOS98,Ohn99,OhM01}, since the magnetization extracted from these
experiments agrees reasonably well with that obtained from direct
SQUID-magnetometer measurements \cite{Ohn99,OhM01} and from magnetic
circular dichroism \cite{Beschoten}. This shows the intimate connection
between carriers and impurity spins. Even more compelling evidence is
obtained from electric field-effect experiments on (In,Mn)As
\cite{OCM00,CYM03} and (Ga,Mn)As \cite{NKS03}, in which the carrier
concentration is altered by the application of a gate voltage in a
field-effect-transistor geometry. The experiments show a very strong
dependence of the coercive force on carrier concentration, besides a
significant dependence of $T_{\mathrm{c}}$.

In (Ga,Mn)As there is a metal-insulator transition between
insulating samples with small manganese concentration and
metallic samples with larger concentration.
Insulating behaviour is here characterized by a diverging
resistivity for $T\to 0$, indicating localization of carriers. Conversely,
in metallic samples the resistivity decreases and eventually saturates for
$T\to 0$. In this paper the term ``metal-insulator transition''
refers to the quantum-phase transition at $T\to
0$ and not to crossovers at finite temperature, as it is sometimes used in
the DMS literature. It is important to note, however, that even for the
most metallic samples the resistivities are relatively high, of the order
of $10^{-3}$ to $10^{-2}\,\Omega\mathrm{cm}$
\cite{MOS98,HHK01,Potashnik,EWC02}. They are thus \emph{bad metals},
showing that disorder is rather strong even in this regime.

We introduce a number of quantities for later use: We denote the
concentration of magnetically active impurities by $n_{\mathrm{m}}$. For
III-V semiconductors with zincblende structure these are assumed to be
substitutional impurities and in the absence of interstitial defects
$n_{\mathrm{m}}$ is related to the doping fraction $x$, \textit{e.g.}, in
$\mathrm{Ga}_{1-x}\mathrm{Mn}_x\mathrm{As}$, by
$n_{\mathrm{m}}=4x/a_{\mathrm{c}}^3$, where $a_{\mathrm{c}}$ is the length
of the conventional cubic unit cell of the fcc cation sublattice. The
\emph{carrier fraction} $p$ is the number of carriers per magnetically
active impurity. This fraction is typically smaller than unity due to
compensation. We refer to the charge carriers as \emph{holes} to make the
presentation more readable but the discussion also holds for electrons in
the case of (largely hypothetical) n-type DMS.

\subsection{Types of impurities}
\label{sus.types}

As noted above, substitutional manganese in GaAs and most other III-V
se\-mi\-con\-duc\-tors (but not in GaN) forms a shallow acceptor state made
up of valence-band states \cite{LJM97,OKR98}. The strong
Coulomb-Hubbard interaction between the electrons in the d-shell, $U
\approx 3.5\:\mathrm{eV}$ \cite{OKM99}, leads to a large Mott-Hubbard
splitting of the d-states. Typically, \textit{ab-initio} calculations tend
to find a significant d-orbital admixture close to the Fermi energy
\cite{Aka98,MaM02,BKS03,KTD03,MaM03,MaZ03a}. This is at least partly due to
the failure of the local-density or local-spin-density approximation to
correctly describe this splitting. The consequence is that partially
filled d-bands are predicted to lie close to the Fermi energy. Recently,
density-functional calculations using improved methods such as LDA$+$U
\cite{PKM00,Kud03,MaZ03a} and self-interaction-corrected LSDA
\cite{SIC,FSS03} show less d-orbital weight in the relevant states,
approaching a picture of valence-band dominated impurity states. In
(Ga,Mn)As both the occupied and the empty d-states apparently end up far
from the Fermi energy. Consequently, charge fluctuations in the d-shell are
suppressed, leaving only the spin degrees of freedom, exchange-coupled to
the carrier spins \cite{ScW66}. Since the d-shell is half filled, Hund's
first rule predicts a local spin $S=5/2$, in agreement with experiments
\cite{LJM97,MOS98,OKR98}.

The shallow-acceptor model for manganese in GaAs is discussed in detail in
\cite{BhB00}. The $112\:\mathrm{meV}$ binding energy of an isolated
manganese impurity level is made up of a larger contribution
($86\:\mathrm{meV}$) due to Coulomb attraction and a smaller contribution
($26\:\mathrm{meV}$) from the exchange interaction with the local manganese
spin \cite{BhB00}.

For the understanding of the spatial distribution of defects in DMS it is
important to check the mobility of substitutional manganese. While not much
is known quantitatively, experiments on digital heterostructures
\cite{KJC00,JSG03} show that part of the substitutional manganese moves
several lattice constants during growth at $230^\circ\mathrm{C}$, without
additional annealing. Vacancies may play a role in the diffusion of
substitutional defects \cite{Potashnik}, lowering the energy barriers for
this diffusion.

For DMS grown by low-temperature MBE a high density of other defects is
expected. Arsenic antisites are known to occur also in pure GaAs grown at
low temperatures \cite{Look,LPN95,Lutz,SNG01} and are also believed to be
present in (Ga,Mn)As \cite{MOS98,Ohn99,SaD03}. In low-temperature growth
under arsenic overpressure, a high concentration of manganese is expected
to lead to increased incorporation of the oppositely charged antisites
\cite{TSO,BKS03,MTD03}. This is supported by experiments
\cite{GND00,SFM01}. The antisite concentration strongly depends on the
growth technique; when $\mathrm{As}_4$ quadrumers are cracked to form
$\mathrm{As}_2$ dimers before they hit the DMS surface, the antisite
concentration can be strongly reduced \cite{EWC02,Schiffer,Gallagher}.
Conversely, in samples grown with a $\mathrm{As}_4$-dominated flux
\cite{MOS98,Ohn99} the concentration of antisites is high and responsible
for most of the compensation. The mobility of antisites is believed to be
low \cite{Sue90}. See, however, \cite{Potashnik} for a different view.
Arsenic \emph{interstitials} may also be present but have received little
attention.

The situation is simpler in Ge$_{1-x}$Mn$_x$ since antisite defects cannot
exist by definition so that compensation is most likely due to interstitial
manganese. Apart from this fact, much of the discussion in this paper
applies to $\mathrm{Ge}_{1-x}\mathrm{Mn}_x$ as well.

The third type of impurity important in manganese-doped DMS is manganese
\emph{interstitials}. The presence of interstitials has been proposed by
Ma\v{s}ek and M\'aca \cite{MaM01} and demonstrated by channelling
Rutherford backscattering experiments by Yu \etal \cite{YWW02}. For a total
manganese concentration corresponding to $x\sim 0.07$ about 17\% of
manganese impurities were found to be in interstitial positions
\cite{YWW02}. Under the typical As-rich growth conditions,
\textit{ab-initio} calculations predict a substantially lower energy of
substitutional manganese defects compared to manganese interstitials
\cite{Erwin,MaZ03}. According to Erwin and Petukhov \cite{Erwin}, a
relatively large concentration of interstitials is nevertheless
incorporated because close to the growing surface manganese positions that
develop into interstitials in the bulk are energetically preferred.

There are two relevant non-equivalent interstitial positions: Manganese may
be in tetrahedral positions coordinated by four arsenic or four gallium
ions (the hexagonal position is much higher in energy \cite{MaM03}). The
arsenic-coordinated T(As$_4$) position is locally similar to the
substitutional position but the next-nearest-neighbor shell is very
different. Intuition suggests that it should be lower in energy than the
gallium-coordinated T(Ga$_4$) position with its short cation-cation bonds
\cite{Gallagher}. \textit{Ab-initio} calculations indicate that it is
indeed lower by $0.3$ to $0.35\:\mathrm{eV}$ assuming charged interstitials
\cite{Gallagher,SE.priv}. Interstitial manganese acts as an double donor.
Also, according to \cite{YWW02,MaM03} subsitutional and interstitial
manganese defects have the tendency to occupy adjacent positions due to
their Coulomb attraction. The interstitial then occupies a T(Ga$_4$)
position. In this situation, the two impurity spins should interact
\emph{antiferromagnetically} due to superexchange. The strength of this
interaction is at least $-26\:\mathrm{meV}$ \cite{MaM03}, leading to the
formation of spin singlets that do not participate in ferromagnetic order.
There is at present no agreement on the exchange interaction between
T(Ga$_4$) manganese spins and carrier spins. According to \cite{BlK03} the
spin of the interstitial is nearly decoupled from the carriers, whereas in
\cite{MaM03} the interaction is found to be only a little smaller than for
substitutional manganese. For the T(As$_4$) position, the interaction is
practically the same as for substitutionals \cite{MaM03},
reflecting the similar neighbourhood. Manganese interstitials are
thought to be highly mobile \cite{YWW02} due to lower energy barriers
for their motion as compared to substitutionals.

\subsection{From isolated impurities to the heavy-doping limit}
\label{sus.crossover}

With increasing concentrations of magnetic impurities and holes there is a
series of crossovers from first isolated impurities with weakly bound holes
to an impurity band, which finally merges with the valence band. We discuss
some important aspects of these crossovers in the following.

The magnetic impurities bind holes in hydrogenic impurity states. Due to
the exchange interaction $J_{\mathrm{pd}}$ between hole spins and defect
spins, the two spins align with each other at sufficiently low
temperatures. In the case of Mn in GaAs, the exchange interaction is
antiferromagnetic \cite{OKR98,Ohn99,BhB00,SOH01}. The impurity-state wave
function falls off roughly at the lenfth scale of the Bohr radius 
$a_{\mathrm{B}}$. To be more precise, the hopping integral between two
ground-state (1s) wave functions reads \cite{Bhatt}
\be
T(\mathrm{R}) = E_0\, \left(3+\frac{3R}{a_{\mathrm{B}}}
  + \frac{R^2}{3a_{\mathrm{B}}^2}\right)\,
  \exp\!\left(-\frac{R}{a_{\mathrm{B}}}\right) ,
\label{TofR.2}
\ee
where $E_0$ is the binding energy of an isolated dopand. A central-cell
correction has been neglected here. The hopping integral thus falls off to
$1/e$ of its value at zero separation on a length scale of
$\overline{a}=2.394\,a_{\mathrm{B}}$. To have only \emph{a single} impurity
in a volume of $\overline{a}^3$, the impurity concentration has to be
relatively low, $1.536 \times 10^{20}\:\mathrm{cm}^{-3}$ corresponding to
$x=0.7$\% in (Ga,Mn)As, outside or
perhaps at the border of the concentration range where magnetic order is
observed. Thus typically a hole spin sees several impurity spins even in
the insulating regime. This conclusion is not changed by spin-orbit
coupling, although it has a significant
effect on the impurity-state wave function and on the hopping integral
\cite{FZD03}.

If there are several acceptors in the range $\overline{a}$, it does not
make sense to keep only the Coulomb attraction of one of them and neglect
the others. In other words, the hole will not be in a hydrogenic impurity
state centered at one acceptor but rather in a more extended
molecular-orbital-like state. This has already been pointed out in
\cite{KS}. Coming from low concentration, the first step is to take dopand
\emph{pairs} into account \cite{BhR81}. A description relying on hydrogenic
impurity wave functions is of limited applicability for $x\gtrsim 0.7$\%.
To make this argument more precise, one should calculate the width of the
impurity band from such a model. The assumption of hydrogenic impurity
states tends to overestimate this width \cite{comment,reply}. Hence, if the
calculated width becomes comparable to the binding energy of the acceptor,
the assumption becomes questionable.

It is energetically favourable for a localized hole spin to align (in
parallel or antiparallel depending on the sign of $J_{\mathrm{pd}}$) with
all the impurity spins in the vicinity, forming a \emph{bound magnetic
polaron} (BMP). This leads to an effective \emph{ferromagnetic} coupling
between impurity spins. If the hole concentration were very small compared
to the concentration of magnetic impurities, \textit{i.e.}, $p\ll 1$, at
low temperatures the system would consist of isolated BMP's and additional
isolated impurity spins in regions with exponentially small hole density
\cite{KaS02,KS,YaM03}. In this regime no magnetic order is expected except
perhaps at exponentially small temperatures but even there it may be
destroyed by quantum fluctuations.

In real samples the hole concentration is smaller than the concentration of
acceptors due to compensation, but typically not by more than one order of
magnitude. Many theoretical papers assume a typical fraction of holes per
manganese impurity in (Ga,Mn)As of 10\% based on earlier experimental
papers \cite{MOS98,Ohn98}, but these estimates have been corrected upwards
\cite{Ohn99} and significantly higher values are now reached in any case
\cite{EWC02}. In the case of (Ga,Mn)As, the concentration of antisites can
be reduced by growing with $\mathrm{As}_2$ dimers
\cite{EWC02,Schiffer,Gallagher}, as noted above. A hole fraction $p$ above
90\% was achieved for an Mn doping level of $x\approx 1.5$\% \cite{EWC02}.

Note that the \emph{Kondo effect} \cite{Kondo,Yosida}
is irrelevant for the DMS studied to date \cite{SHK03}. Kondo physics is
relevant in the regime of large carrier concentration compared to the
impurity concentration, $p\gg 1$. DMS are so far always in the opposite
regime of $p<1$. However, in principle the DMS and Kondo regimes are
continuously connected and it would be interesting to study the crossover
to Kondo physics by increasing the carrier concetration, \emph{e.g.}, by
codoping or even by application of a gate voltage.

For increasing impurity and hole concentrations the typical separation of
holes eventually becomes of the order of $\overline{a}$. This happens for
$p n_{\mathrm{m}} \overline{a}^3 \sim 1$, \textit{i.e.}, $p x \sim 0.7$\%
for (Ga,Mn)As. For strong compensation ($p\ll 1$) the corresponding
impurity doping $x$ is rather large. At this doping level the typical BMP's
start to overlap. For higher concentrations a typical impurity spin has
appreciable interaction with a number of hole spins. Of course, this
interpretation only holds if it still makes sense to talk about carriers in
individual hydrogenic impurity states. This can be checked using a theory
that can describe both localized and extended states, see below.

In the band picture, isolated impurities are characterized by impurity
levels. As the concentration of impurities is increased, some impurity
states start to overlap. First, new impurity levels corresponding to close
\emph{pairs} of impurities will appear at energies depending on their
separation \cite{BhR81}. At higher concentrations of magnetic impurities,
these states merge into an impurity band. The width of this band increases
with the impurity concentration. Most of the states in the impurity band
are localized due to the disordered impurity positions. Eventually the
impurity band merges with the valence band and looses its independent
character. In this heavy-doping limit the attractive Coulomb potential of
the dopands is often neglected in the literature. Then the DMS is described
in terms of holes in the valence band, which only interact with the
impurities through their exchange interaction. This physics is described by
the Zener model \cite{Zener,Zener1,DHM97,DOM00} discussed in
\sref{sus.trans.band}.

However, there is no need to ignore the Coulomb interaction with the
defects in a model starting from holes doped into the valence band. An
acceptor introduces a hole and provides an attractive Coulomb potential. If
the impurity level is shallow the dopands can be well described in an
envelope-function formalism \cite{KSJ03}. To obtain the correct numerical
value of the binding energy one has to include a phenomenological
central-cell correction \cite{BhB00}, which models the stronger Coulomb
attraction of the hole to the impurity nucleus at short distances. Such a
model reproduces the isolated bound impurity state but is applicable in the
\emph{entire} doping range up to heavy doping. It can thus, in principle,
be employed to decide when (a) the picture of weakly interaction hydrogenic
impurity states and (b) the neglection of the Coulomb disorder potential
for valence-band holes are justified. Such an approach is applicable if
there is no strong admixture of d-orbitals of the magnetic impurity to the
hole states \cite{LJM97,OKR98}.

\subsection{CPA vs.\ supercell calculations}
\label{sus.CPA}

Finally we discuss possible ways to include impurities in theories of DMS.
There are two basic approaches: First, one can consider a relatively large
section of the lattice with many impurities and periodic boundary
conditions. If this \emph{supercell} is sufficiently large, one can capture
disorder effects such as weak localization. This approach allows to
investigate the dependence of observables on system size, which allows to
use finite-size scaling to determine the localization length of states. It
also allows to treat not only fully random disorder but also impurities
with correlated positions. We will see in \sref{sec.prop} that such
correlations appear naturally in DMS. The supercell approach can be used,
in principle, in both \textit{ab-initio} and model theory but quickly
becomes very costly for \textit{ab-initio} calculations. For this reason it
has mostly been applied to a supercell with a single substitutional
manganese impurity; a concentration of $x=0.0625$ already requires a
supercell of 32 atoms. At present, supercells large enough to capture
disorder effects do not seem feasible in \textit{ab-initio} approaches.

The second approach consists of the dynamical coherent-potential
approximation (CPA), which treats the carriers as independent particles in
an effective medium \cite{VEK68,Kub74,Fau82,RAS95,TaM96}. In the dynamical
CPA this medium is described by a spin-dependent \emph{purely local}
self-energy $\Sigma_\sigma(\omega)$, which is determined from the CPA
condition that the averaged carrier $T$-matrix vanishes \cite{TaM96}. The
average here consists of both a spatial average over all sites, taking into
account that some of them are occupied by impurities, and a thermal average
over the orientations of impurity spins. The CPA includes multiple elastic
scattering off a single impurity but not inelastic
scattering. The CPA does not describe localization
effects and is thus limited to the metallic regime. Since it is an
effective-medium approach it cannot be used to study effects of the spatial
distribution of defects. It is difficult to treat scattering potentials in
CPA that are not purely local. This has been done in other condensed-matter
systems \cite{AHM00} but not, to the author's knowledge, in DMS. On the
other hand, the CPA allows to change the doping level continuously and does
not require the use of a large supercell. It is thus also suited for
\textit{ab-initio} calculations. A comparison of \textit{ab-initio} results
obtained with the CPA and for a supercell is given in \cite{BKS03}.

\section{Properties of disorder}
\label{sec.prop}

Due to the relatively low carrier concentration, electronic screening is
rather weak and Coulomb interactions are correspondingly strong. From
nonlinear screening theory \cite{SE} one obtains a screening length of
\be
r_{\mathrm{s}} = \frac{(3n_{\mathrm{m}}-2n_{\mathrm{h}})^{1/3}}
  {n_{\mathrm{h}}^{2/3}} = \left(\frac{3-2p}{4xp^2}\right)^{\!1/3}\,
  a_{\mathrm{c}} .
\ee
In (Ga,Mn)As, where $a_{\mathrm{c}}=5.653\:\mbox{\AA}$, the screening
length is $r_{\mathrm{s}}=108.1\:\mbox{\AA}$ for $x=0.01$, $p=0.1$ and
$r_{\mathrm{s}}=28.9\:\mbox{\AA}$ for $x=0.05$, $p=0.3$. In the weak-doping
limit the nonlinear screening theory becomes inapplicable, though
\cite{SE}. The screening length is thus larger than the typical length
scale $\overline{a}$ of the hopping integral and the separation between
impurities. Also, because of the $x^{-1/3}$ dependence of $r_{\mathrm{s}}$
we expect $r_{\mathrm{s}}$ to remain larger than this separation, as long
as nonlinear screening theory remains valid. Furthermore, due to
compensation, typically many defects of either charge are present.
Consequently, neighbouring defects typically experience a nearly
unscreened Coulomb interaction.

Let us first turn to the Coulomb disorder potential. Typically several
charged defects are present within a sphere of radius $r_{\mathrm{s}}$.
Most papers that consider the microscopic defect positions at all assume a
random distribution. However, the weakly screened Coulomb interaction makes
a random distribution of defects very costly in energy. We expect two
effects that reduce the Coulomb energy: Firstly, during growth impurities
are not incorporated randomly but in partially correlated positions, in
particular due to the attraction of oppositely charged defects
\cite{BKS03,MTD03}. Secondly, defect diffusion \cite{Potashnik} leads to a
rearrangement of defects that is energetically more favourable. In
\cite{TSO} the resulting impurity distributions are studied in the extreme
case that the defects approach thermal equilibrium with respect to the
Coulomb interaction. In \cite{icps,pasps} the approach towards equilibrium
during annealing is also considered. The compensation is assumed to be due
to antisites. In the following, these results are briefly reviewed and a
number of new results are presented.

In \cite{TSO} Monte Carlo (MC) simulation of charged manganese
substitutionals and arsenic antisites in (Ga,Mn)As at a typical growth and
annealing temperature of $250^\circ\mathrm{C}$ is used to obtain
configurations close to equilibrium. The holes enter through the nonlinear
screening length $r_{\mathrm{s}}$ \cite{SE}. The resulting equilibrium
distribution is assumed to be quenched at low temperatures. The density of
antisites is determined by charge neutrality from the manganese and hole
concentrations estimated from experiment \cite{MOS98,Ohn99,EWC02}. The main
result is that for all realistic manganese and antisite concentrations the
impurities arrange themselves in \emph{clusters}, which typically contain
oppositely charged manganese and antisite defects at nearest-neighbour
positions on the cation sublattice. The cluster formation leads to a
reduction of the Coulomb potential. For example, an antisite defect (charge
$+2$) with two manganese impurities (charge $-1$) at nearest-neighbour
positions is electrically neutral and only has a dipole or quadrupole field
at large distances. Interestingly, Lee \etal \cite{LJM02} suggest that
\emph{intentional} clustering of magnetic impurities will be used to
produce functional heterostructures of DMS. Note that a short-range
attraction between substitutional manganese has been found in
\textit{ab-initio} calculations \cite{ScM01,RAN03}. This chemical effect is
not included in \cite{TSO}, but it would be easy to modify the potential
accordingly.

Scanning tunneling microscopy for (Ga,Mn)As with low doping, $x=0.6$\%,
does not show clear signs of clusters \cite{GND00,SBW03}. However, this
method is only sensitive for impurities within the topmost two layers. On
the other hand, the experiments do show a significantly larger abundance of
features attributed to defect pairs than expected for a random distribution
\cite{SBW03}. It would be interesting to repeat these experiments for
samples with large $x$.

\begin{figure}
\centerline{\includegraphics[width=9.5cm]{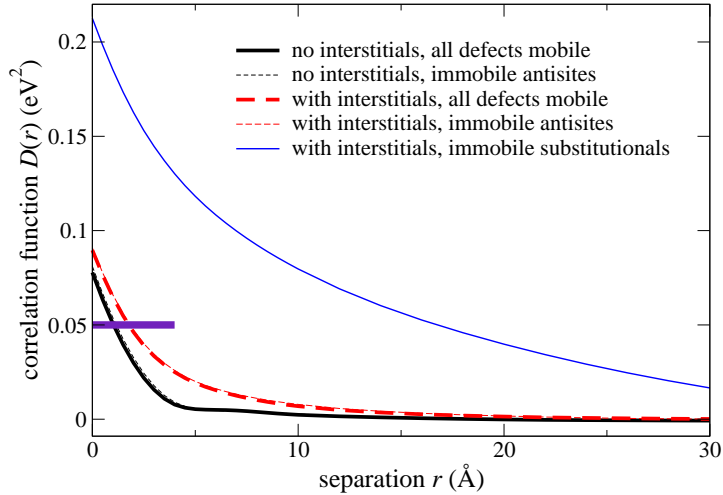}}
\caption{\label{fig.Dofr}Correlation function $D(r)$ of the Coulomb disorder
potential. All curves have been obtained performing sufficiently many MC
steps to reach convergence, starting from a random distribution of
defects at the appropriate sites in a supercell of $20\times
20\times 20$ conventional fcc unit cells (32000 cations). Substitution of a
fraction of $x=0.05$ of cation sites (1600) by manganese and $p=0.3$
holes per substitutional manganese ion have been assumed in all cases. 
For the curves marked ``no interstitials,'' compensation is entirely due to
antisites (560 in this system) as in \protect\cite{TSO,icps,pasps},
whereas for the curves marked ``with interstitials,'' 20\% of all
manganese is in interstitial positions (corresponding to 400 interstitials
and 160 antisites).
The broad horizontal bar denotes the nearest neighbor separation on the
cation sublattice ($3.9973\:$\AA).}
\end{figure}

Quantitative information on the Coulomb disorder can be obtained from the
correlation function
\be
D(r) \equiv \langle
   V(\bi{r})\,V(\bi{r}')
   \rangle_{\scriptscriptstyle |\bi{r}-\bi{r}'|=r} -
   \langle V \rangle^2
\ee
of the Coulomb disorder potential $V(\bi{r})$. Obviously, $\Delta
V\equiv\sqrt{D(0)}$ is the width of the distribution of $V(\bi{r})$. Defect
clustering has two effects: It strongly reduces $D(r)$ (and thus $\Delta
V$) and it makes $D(r)$ short-range correlated. While the reduction of
$\Delta V$ is substantial, $\Delta V$ is still \emph{not} small compared to
the Fermi energy \cite{TSO} even in the heavy-doping regime so that
disorder cannot be neglected. As shown in \fref{fig.Dofr},
in equilibrium $D(r)$ decays to only about
10\% of $D(0)$  on the scale of the nearest-neighbor separation. Thus
$D(r)$ is well approximated by a delta function. This initial decay is due
to the screening of the compensated manganese impurities \cite{icps,pasps}.
Clearly, \emph{ionic screening} by charged impurities is nearly perfect.
The remaining uncompensated manganese ions cannot be screened by
antisites and their contribution decays on a
larger length scale determined by their density.

In (Ga,Mn)As antisite defects are believed to be immobile at typical growth
and annealing temperatures \cite{Sue90}. However, the mobility of antisites
in nearly perfect GaAs may not be a good indicator for the mobility in
heavily doped (Ga,Mn)As, for which there is often another impurity in the
nearest-neighbour shell. It would be worthwhile to study the mobility of
antisites and substitutional manganese experimentally. In any case, if the
antisites are held fixed in the MC simulations,  the correlation function
$D(r)$ is practically unchanged, see \fref{fig.Dofr}. Since there are much
more manganese impurities than antisites, the manganese ions can easily
screen immobile antisites.

Fiete \etal \cite{FZD03} also perform MC simulations for the positions of
substitutional manganese impurities in weakly doped (Ga,Mn)As. The
objective is to introduce some degree of correlations between defect
positions, similarly to \cite{TSO}. For large numbers of MC steps a nearly
perfect body-centered cubic lattice of impurities is found. This is not
surprising since no compensating defects are included, \textit{i.e.}, all
impurities have the same charge. Compensating defects are necessary for
cluster formation \cite{TSO}.

The next step necessary for a realistic description is the incorporation of
manganese \emph{interstitials} \cite{YWW02}. We have performed MC
simulations similar to \cite{TSO} including charged ($+2$) interstitials
and assuming the onsite energy in the T(Ga$_4$) position to be
$0.3\:\mathrm{eV}$ higher than in the T(As$_4$) position
\cite{Gallagher,SE.priv}. \Fref{fig.Dofr} shows that the width $\Delta V$
as well as the correlation length increase slightly due to interstitials,
if substitutional manganese is assumed to be mobile, regardless of the
mobility of antisites. This is partly due to the fact that
the MC simulation minimizes the configuration energy including the
different on-site energies of interstitials, whereas $D(r)$ is the
correlation function of the Coulomb potential alone. The correlation length
is still of the order of the nearest-neighbor separation, though.

Assuming \emph{only} the interstitials to be mobile---probably an
unrealistic assumption \cite{KJC00,JSG03}, even though interstitials are
the most mobile defects---and starting from a fully random distribution of
substitutional manganese impurities, the interstitials move to positions
close to two or more substitutionals. In equilibrium, most reside in
T(As$_4$) positions due to the lower onsite energy. The T(Ga$_4$) sites are
closer to the substitutionals and thus reduce the Coulomb energy, but this
gain is overcompensated by the higher onsite energy. The correlation
function in \fref{fig.Dofr} decays much more slowly, similarly to the case
of random defects. The reason is that the high concentration of
substitutional manganese impurities cannot be effectively screened by the
lower concentration of interstitials. Of course, results like this depend
on details of the configuration energy and are to be interpreted with
caution.

In \cite{MaZ03}, small clusters of three manganese defects are studied with
a density-functional total-energy method. Clusters made up of two
substitutional and one interstitial defects are found to be favoured due to
their Coulomb attraction, in qualitative agreement with the above
discussion. Of course, \textit{ab-initio} theory captures additional
``chemical'' contributions to the energy not included in \cite{TSO}.

\section{Effects of disorder on transport}
\label{sec.trans}

In the present section the effect of disorder on electronic transport is
discussed. Experimentally, the resistivity for $T\to 0$ is found to either
saturate (metallic behaviour) or diverge (insulating behaviour). As
discussed above, metallic behaviour is seen for higher concentrations of
magnetic impurities and thus, more relevantly, higher carrier
concentrations. A metal-insulator transition is observed as a function of
impurity concentration \cite{MOS98,Ohn99}. Annealing at low temperatures
for short times is found to make the samples more metallic
\cite{Potashnik}, whereas longer annealing \cite{Potashnik} or annealing at
higher temperatures \cite{EBB97} has the opposite effect. This is consistent
with the expectation that higher disorder makes the sample less metallic.

At higher temperatures, the resistivity shows a clear peak centered around
$T_{\mathrm{c}}$ \cite{EBB97,MOS98,Ohn99,HHK01,EWC02}. Only for samples far
in the insulating regime, where the resistivity increases very strongly for
$T\to 0$, this peak develops into a shoulder \cite{MOS98}. This feature is
very robust; it even survives in ion-implanted samples, where the disorder
is very strong and dominated by implantation damage \cite{PLS03}.

The Fisher-Langer theory \cite{FL} relates the fluctuation corrections of
the resistivity of ferromagnets to their magnetic susceptibility. For a
susceptibility of Ornstein-Zernicke form it predicts a infinite
\emph{derivative} of $\rho$ at $T_c$ and a very broad maximum at higher
temperatures \cite{FL} in good agreement with experiments for ferromagnetic
\emph{metals} but not for DMS. To apply the theory \cite{FL} to DMS, the
magnetic susceptibility at large $\q$ should be obtained \cite{sus}.

The physics of impurities discussed in \sref{sus.types}
is expressed by a Hamiltonian of the form
\be
H = H_{\mathrm{b}} + H_{\mathrm{m}} - J_{\mathrm{pd}} \sum_n
  \bi{S}_i\cdot \bi{s}(\bi{R}_n) ,
\label{H.fullx.3}
\ee
where $H_{\mathrm{b}}$ and $H_{\mathrm{m}}$ contain terms concerning only
carriers and impurity spins, respectively, and the last term describes the
local exchange coupling between carrier and impurity spins. In principle,
$H_{\mathrm{b}}$ contains not only the band structure of the host
semiconductor but also the electron-electron interaction and the
interaction of electrons with the disorder potential due to impurities. The
electron-electron interaction is assumed to be partly incorporated in the
band structure, \textit{e.g.}, on the Hartree-Fock level, and the remaining
correlation terms are usually neglected in the DMS literature (they might
be important, though \cite{YaM03}). For very strong doping the Fermi energy
measured from the band edge is large compared to the width of the disorder
potential. In this limit, Coulomb disorder from the impurities may be
ignored. However, even for 5\% manganese in (Ga,Mn)As this condition is
not satisfied \cite{TSO,JAS02} so that the omission of Coulomb disorder is
a rather severe approximation.

While the Coulomb disorder potential is most important for the observed
metal-insulator transition, the additional disorder due to the exchange
interaction with impurity spins shifts the transition further towards
higher carrier concentrations \cite{DMO03}. By aligning the impurity spins
in an external magnetic field, this source of disorder can be reduced,
leading to less scattering for higher fields and thus to a negative
magnetoresistance \cite{DMO03}. The system can even be driven to the
metallic side of the transition by an external field \cite{KOI98,FCW01}.

\subsection{Band picture}
\label{sus.trans.band}

Unperturbed host carriers (zero disorder potential) with an exchange
interaction with impurity spins are described by the \emph{Zener model}
\cite{Zener,Zener1,DHM97,DOM00}. This model still contains disorder since
the positions of impurity spins appear in equation \eref{H.fullx.3}. To
obtain an analytically tractable theory, a \emph{virtual crystal
approximation} (VCA) is often made, whereby the discrete impurity spins are
replaced by a smooth spin density \cite{DHM97,JAL99,DOM00,KSJ03}. This may
be valid since the Fermi wavelength $\lambda_{\mathrm{F}}$ is typically
larger than the separation between impurity spins and a carrier-mediated
interaction automatically averages over length scales of the order of
$\lambda_{\mathrm{F}}$.

A simple way to restore part of the effects of disorder in the VCA is to
assume a finite quasiparticle lifetime \cite{SJY02,JAS02,SJM03,JSW03}. The
Coulomb and exchange contributions to this lifetime are estimated in
\cite{JAS02}. The dominant Coulomb scattering leads to a quasiparticle
width $\Gamma$ of the order of $150\:\mathrm{meV}$. This approach is only
applicable in the strongly metallic regime since it does not capture
localization. The effect of $\Gamma>0$ is clearly seen in theoretical
results for the optical conductivity, where it leads to a broadening of the
intraband (Drude) and interband peaks \cite{SJY02}.

A related approach to the incorporation of disorder is employed in
\cite{sus}, where semiclassical Boltzmann equations are used to describe
the dynamics of carriers and impurity spins paying special attention to
spin-orbit effects. The correlation function $D(r)$ of the Coulomb disorder
potential can be assumed to be delta-function-correlated \cite{TSO}, see
\sref{sec.prop}. If $D(r)\approx \gamma\delta(\bi{r}-\bi{r}')$, the leading
self-energy diagram yields the scattering rate $\hbar/\tau = 2\pi\gamma
N(E_{\mathrm{F}})$ \cite{LeR85}, where $N(E_{\mathrm{F}})$ is the
electronic density of states per spin direction. The delta-function
correlation also simplifies the description of scattering by collision
integrals in the Boltzmann formalism. This approach yields analytical
expressions for the carrier and impurity spin susceptibilities depending on
the scattering rates from Coulomb and exchange disorder \cite{sus}.

We now turn to approaches that treat the disorder potentials explicitly.
Taking the Coulomb disorder potential into account,
Timm \etal \cite{TSO,icps,pasps}
study the hole states in the Coulomb disorder potential. The additional
disorder due to the exchange interaction is small compared to the Coulomb
disorder \cite{JAS02}. The impurity distribution is obtained from MC 
simulations. The envelope function and parabolic-band approximations are
employed. For quantitative calculations the detailed band
structure should be taken into account, \emph{e.g.}, using the 6-band
Kohn-Luttinger Hamiltonian \cite{KL,AJB01,DOM01,SKM01}.
The hole Hamiltonian is written
in a plane-wave basis and diagonalized numerically, giving the energy
spectrum and eigenfunctions $\psi_n({\bf r})$. The normalized eigenfunctions
determine the participation ratios \cite{EdT72}
\be
\mathrm{PR}(n)=\left(\sum_{\bf r} |\psi_n({\bf r})|^4\right)^{\!-1}
\ee
of the states. (In part of the literature, including \cite{TSO,icps,pasps},
this quantity is called ``inverse participation ratio'' but it seems more
natural to follow \cite{EdT72} and call it ``participation ratio.'')
The participation ratio quantifies the number of lattice sites
where the state $n$ has a significant probability. Hence,
the participation ratio scales with the
system size for extended states but essentially remains constant for
localized states. It thus allows to estimate the position of the
mobility edge in the valence band.

\begin{figure}[ht]
\centerline{\includegraphics[width=8cm,clip]{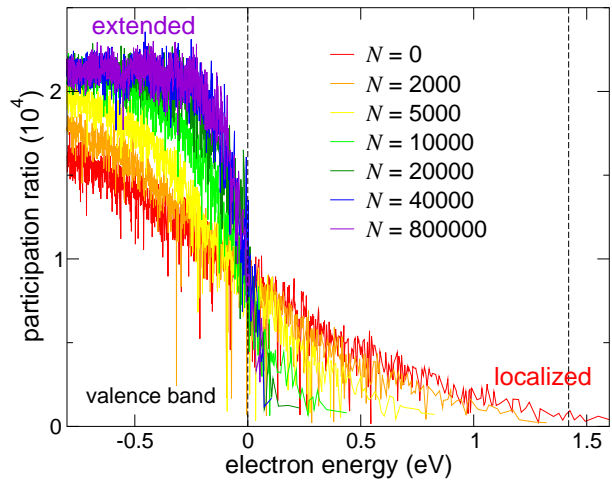}}
\caption{\label{fig.pr1}Participation ratio as a function of
electron energy for $x=0.05$ and $p=0.3$
after various numbers $N$ of MC steps increasing from the
flattest to the steepest curve, after \protect\cite{icps}.
Zero energy is at the unperturbed valence-band edge.}
\end{figure}

\Fref{fig.pr1} shows the participation ratio as a function of electron
energy for $x=0.05$ and $p=0.3$, after various numbers of MC steps $N$
\cite{icps}. The plot shows that for random defects ($N=0$) the band gap is
filled in due to disorder, which is in contradiction to experiments. Thus
correlated defects are required to explain the persistence of the energy
gap. Finite-size scaling shows that the participation ratio in the flat
region and the upper part of the slope scales with system size, whereas it
become independent of system size in the band tail \cite{icps}. The
mobility edge thus lies on the slope in \fref{fig.pr1}. Since the MC 
simulation and also the diagonalization become numerically demanding for
large system sizes, the finite-size scaling could not be carried out for
sufficiently large systems to pinpoint the mobility edge exactly.

\begin{figure}[ht]
\centerline{\includegraphics[width=9cm,clip]{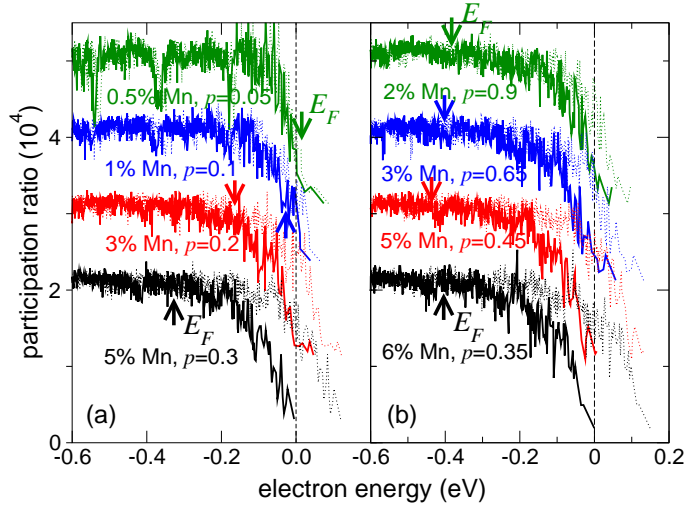}}
\caption{\label{fig.pr2}Participation ratio as a function of electron energy
for various parameters $x$ and $p$ given in the plot, (a) for the (Ga,Mn)As
samples of \protect\cite{MOS98,Ohn99}, after \protect\cite{TSO}, (b) for the
samples of \protect\cite{EWC02}. All impurity configurations have
relaxed to thermal equilibrium in the MC simulation. For the
exchange splitting full magnetization of manganese spins has been assumed.
The Fermi energy is in each case indicated by an arrow.}
\end{figure}

For the transport properties the states close to the Fermi energy are the
most important. To find the position of the Fermi energy one has to take
into account the splitting of the valence band by the exchange interaction.
While the \emph{disorder} due to this interaction is small, its
\emph{average} is not. The result is that for random defects the Fermi
energy lies on the slope of the curve in \fref{fig.pr1}, showing a strong
tendency towards localization \cite{TSO}, whereas the states at the Fermi
energy are extended for the ``annealed'' (clustered) configuration. Thus in
this model cluster formation is required to understand why (Ga,Mn)As with
$x=0.05$ is metallic. \Fref{fig.pr2} shows the participation ratio as a
function of energy for parameters appropriate for (a) the samples meassured
by Matsukura, Ohno \etal \cite{MOS98,Ohn99} and (b) samples by Edmonds
\etal \cite{EWC02}. It is obvious that the shape of the curves does not
change strongly, due to the strong ionic screening. In (a) the main effect
is the shift of the Fermi energy due to the strongly changing hole
concentration. The states at the Fermi energy are clearly extended for
$x=0.03$ and $x=0.05$, while they are localized for 0.5\% manganese
\cite{TSO}. This is in reasonable agreement with the experimentally
observed metal-insulator transition \cite{MOS98,Ohn99}. On the other hand,
in (b) the Fermi energy hardly changes at all and stays well in the
extended region, consistent with the experimental observation of metallic
transport from $x=1.65$\% to $x=8.8$\% \cite{EWC02,BG.priv}.

Yang and MacDonald \cite{YaM03} extend this picture by not only including
the Coulomb disorder potential due to substitutional manganese and antisites
but also the Coulomb interaction \emph{between holes}. Employing the
Hartree-Fock approximation for the latter and considering the participation
ratios, they also find metallic transport at $x=0.05$ but insulating behaviour at
$x=0.0125$.

One can learn more by controlling the defect distribution by growing DMS
layer by layer. For example, in \cite{KJC00,JSG03} digital heterostructures
consisting of half monolayers of MnAs separated by 10 or 20 monolayers of
GaAs are grown. The carrier concentration is systematically varied by
codoping the GaAs layers with beryllium or silicon. All samples are
insulating for $T\to0$, consistent with strong disorder \cite{JSG03}. There
are indications that part of the manganese diffuses out of the MnAs layers.
The resistivity increases (decreases) if the hole concentration is reduced
(inceased) by codoping \cite{JSG03}. We have performed MC simulations of
the type of \cite{TSO} for a superlattice of manganese-rich layers in GaAs.
The number of MC steps has been chosen such that some interdiffusion occurs
but the manganese-rich layers remain clearly defined. Diagonalization of
the hole Hamiltonian than shows very strong smearing of the valence-band
edge, more so than for a random distribution, and very small participation
ratios (pronounced localization) of most states, in qualitative agreement
with experiments \cite{JSG03}.

Starting from a single-band model \cite{AMD02}, Alvarez and Dagotto
\cite{AlD03} calculate the density of states, optical conductivity and
(magneto-) resistance of DMS. The method \cite{AMD02,AlD03} relies on MC
simulations for the orientations of classical impurity spins interacting
with the holes. The model assumes a simple cubic lattice. The holes can
visit any site but only a fraction of randomly selected sites carries
impurity spins. Defect clusters are not studied. The
Coulomb fields from charged impurities are neglected. This approximation is
fairly restrictive since it removes the Coulomb disorder, only leaving the
weaker disorder from the exchange interaction \cite{JAS02}. The hole
Hamiltonian is diagonalized for each configuration of impurity spins during
the simulation, which restricts the size of the supercell to at most $8^3$
sites \cite{AMD02,AlD03}.

\begin{figure}
\centerline{\includegraphics[width=5cm]{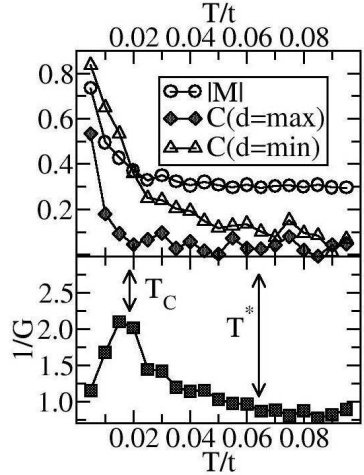}}
\caption{\label{fig.AD}Upper panel: Magnetization $|M|$ and spin-spin
correlation function $C$ from Monte Carlo simulations for a $12\times 12$
lattice with 22 spins ($x\approx 0.15$) and 6 carriers ($p\approx 0.3$) and
$J_{\mathrm{pd}}/t=1$, where $t$ is the hopping amplitude in the
tight-binding model. Lower panel: Inverse of the conductance for the same
model. From \protect\cite{AlD03}.}
\end{figure}

For large $J_{\mathrm{pd}}$ the exchange interaction becomes strong enough
to lead to weak localization of the holes \cite{AMD02,AlD03}. In the band
picture, $J_{\mathrm{pd}}$ leads to the formation of impurity levels which
form an impurity band as the impurity states start to overlap. This is
clearly seen in the density of states and, consequently, in the optical
conductivity calculated in \cite{AlD03}. The DC resistivity is also
evaluated in \cite{AlD03} using the Kubo formula. It shows metallic
behaviour for small $J_{\mathrm{pd}}$ (weak disorder) and insulating
behaviour for large $J_{\mathrm{pd}}$ (strong disorder). Interestingly, as
seen in \fref{fig.AD} (lower panel), for intermediate disorder strength the
resistivity first increases with temperature up to $T\sim T_{\mathrm{c}}$
and then decreases again \cite{AlD03}, very similar to the experimental
results. However, the model does not represent the situation realized in
(Ga,Mn)As and probably most other DMS, where the Coulomb interaction is the
dominant origin of localization. Nevertheless the qualitative results
should be valid regardless of the specific origin of localization.

The same model is studied within the CPA by Bouzerar \etal
\cite{BoP02,BKB02}. The Coulomb potential is omitted but a large exchange
coupling $J_{\mathrm{pd}}$ leads to the formation of an impurity band
\cite{BKB02}, consistent with \cite{AMD02,AlD03}. For intermediate
$J_{\mathrm{pd}}$ this band merges with the valence band.

\subsection{Percolation picture}

Kaminski and Das Sarma \cite{KS} consider the effect of disorder in the
\emph{localized} regime, conceptually starting from the light-doping limit.
Similarly to other works \cite{BB,BBlong,MAD02,Erwin,FZD03} the main
ingredients are holes in hydrogenic impurity states bound to magnetic
acceptors. For \emph{randomly} distributed impurities one obtains systems
of overlapping BMP's \cite{KaS02,KS}, as discussed in \sref{sus.crossover}.
(We reserve the term ``clusters'' to close groups of impurities.) At
sufficiently low temperatures the impurity spins belonging to the same
system are ferromagnetically aligned. If such a domain becomes infinite,
long-range ferromagnetic order ensues. This is discussed further in
\sref{sec.mag}. For strongly localized holes \cite{KaS02}, when the
separation between acceptors is larger than the size of the impurity-state
wave function, transport is due to thermally activated hopping of holes.
Thus the resistivity diverges for $T\to 0$, showing insulating behaviour.
On the other hand, no special signature is expected in the resistivity at
the Curie temperature \cite{KS}, essentially because \emph{all} holes
contribute to the resistivity, not only those in the infinite ferromagnetic
domain forming at $T_{\mathrm{c}}$. For $T\lesssim T_{\mathrm{c}}$ this
domain only comprises a small fraction of the sample volume and only for
holes moving within this domain the activation energy is reduced by
ferromagnetic alignment \cite{MAD02,KS}. In addition, the contribution of
the exchange interaction to the hopping activation energy is small compared
to the Coulomb interaction. As discussed in \sref{sus.crossover}, the
picture of strongly localized holes only applies for small doping levels.
While the numerical results rely on the presence of \emph{hydrogenic}
impurity states, the conceptual picture also applies to impurity states
more strongly localized due to d-orbital admixture.

Fiete \etal \cite{FZD03} derive the parameters of a BMP model from a
Kohn-Luttinger Hamiltonian restricted to heavy and light holes in the
spherical approximation, \textit{i.e.}, all band energies depend only on
the modulus of the wave vector, a model introduced earlier in \cite{ZaJ02}.
Due to spin-orbit coupling, the impurity-state wave function is not
spherical \cite{FZD03}. The overlap between impurity states leads to the
formation of an impurity band with a roughly symmetrical density of states.
Employing finite-size scaling, Fiete \etal \cite{FZD03} show that the
states in the band tails are localized whereas they are extended in most of
the band. For a manganese concentration of $x=0.01$ the Fermi energy is
found to lie just on the extended side of the mobility edge. The
metal-insulator transition is concluded to take place in the impurity band
\cite{FZD03}.

How does this picture change if the magnetic impurities form
\emph{clusters} \cite{KS}? If several holes are present in the same cluster
the total energy is increased due to their Coulomb repulsion. For an
isolated impurity state this repulsion should be of the same order as the
impurity-state binding energy so that double occupancy can be ignored. For
a sufficiently large cluster, on the other hand, the charging energy
becomes low enough to allow more than one hole to be present \cite{KS}. The
system is still insulating since the resistivity is now controlled by the
small hopping amplitudes \emph{between} clusters. However, now a maximum in
the resistivity is expected close to $T_{\mathrm{c}}$ \cite{KS}. Its origin
is that now the clusters are ferromagnetically polarized well above
$T_{\mathrm{c}}$, which is also determined by the small hole hopping
amplitudes between clusters. Thus at $T_{\mathrm{c}}$ already a significant
volume fraction becomes ferromagnetically aligned, unlike in the previous
case. Then a significant fraction of the holes experience a smaller
activation energy due to the (nearly) optimal alignment of impurity spins
\cite{KS}. Also, the exchange interaction can be expected to play a larger
role now, since the Coulomb interaction is reduced for the more extended
wave functions. These effects may lead to the observed decrease of the
resistivity below $T_{\mathrm{c}}$.

\section{Effects of disorder on magnetic properties}
\label{sec.mag}

Experimentally, an anomalous shape of the magnetization curve $M(T)$ is
observed in DMS
\cite{MOM89,OMP92,EBB97,Tan98,Beschoten,OhM01,Potashnik,GeMn,MSS02,%
Schiffer,LKF03,PLS03}. From measurements of the temperature-dependent
magnetization and comparison with various theories one can hope to learn
more about DMS and the role of disorder in particular \cite{SHK03}. In
insulating samples the magnetization curve is often \emph{concave} over a
broad temperature range \cite{Beschoten,GeMn,MSS02,LKF03,PLS03},
\textit{i.e.}, $d^2M/dT^2>0$. In metallic samples the magnetization curve
is usually nearly linear over a broad range \cite{Beschoten,OhM01,MSS02}.
These results are in striking contrast to the Brillouin-function-like
behaviour predicted by the mean-field theories of Weiss and Stoner
\cite{Yosida} and observed in most insulating and metallic ferromagnets,
respectively. The origin of the anomalous $M(T)$ is most likely that
disorder is much more important in DMS. In some samples the deviation from
Brillouin-function-like behaviour is less pronounced
\cite{Potashnik,Schiffer}, presumably related to sample quality. The
magnetization measured in an applied field of the order of or larger than
the coercive field typically also follows a Brillouin function
\cite{EWC02,SSS02,BlW02,BlW03}. Due to the applied field, fluctuations are
suppressed, leading to a more mean-field-like curve.

\subsection{Band picture}
\label{sus.mag.band}

We return to the Hamiltonian in equation \eref{H.fullx.3} describing
carriers and impurities coupled by a local exchange interaction.
Progress has been made employing the VCA and decoupling the
exchange interaction at the mean-field level. Several authors
\cite{DHM97,JAL99,DOM00} have rederived variants of the expression of
Abrikosov and Gor'kov \cite{AbG62} for the mean-field Curie temperature,
\be
k_{\mathrm{B}}T_{\mathrm{c}} = \frac{S(S+1)}{3}\,
  \frac{J_{\mathrm{pd}}^2n_{\mathrm{m}}\chi}{g^2\mu_{\mathrm{B}}^2} ,
\label{Tc.4}
\ee
where $n_{\mathrm{m}}$ is the impurity density, $\chi$ is the carrier spin
susceptibility, $g$ is the carrier g-factor and $\mu_{\mathrm{B}}$ is the
Bohr magneton. For a single parabolic band, an approximation more
appropriate for the conduction band than for the valence band, $\chi$
is just the Pauli susceptibility \cite{Yosida}
$\chi_{\mathrm{P}}=N(E_{\mathrm{F}})\,g^2\mu_{\mathrm{B}}/2$,
where $N(E_{\mathrm{F}})$ is the density of states per spin direction at the
Fermi energy.
Then the Curie temperature becomes $k_{\mathrm{B}}T_{\mathrm{c}}
=S(S+1)\,N(E_{\mathrm{F}})\,J_{\mathrm{pd}}^2n_{\mathrm{m}}/6$ \cite{SHK03}.
It depends on the atomic fraction $x$ of substitutional magnetic impurities
and on the fraction of carriers per impurity as $T_{\mathrm{c}} \propto p^{1/3}
x^{4/3}$. Thus the increase of $x$ should lead to
higher Curie temperatures \cite{DOM00,DOM01}. However,
growth of DMS is typically limited to small $x$.

The term $H_{\mathrm{m}}$ in equation \eref{H.fullx.3} contains a short-range
antiferromagnetic su\-per\-ex\-change interaction \cite{And59} between the
impurity spins. In DMS this interaction is small compared to the
ferromagnetic interaction, especially in (Ga,Mn)As
\cite{DOM00,DiO01,DOM01}. Taking into account (a) the ferromagnetic
Stoner-type interaction between the carriers, which increases the Curie
temperature, and (b) the superexchange, which reduces $T_{\mathrm{c}}$ one
obtains the expression \cite{DOM00,DiO01,DOM01}
\be
k_{\mathrm{B}}T_{\mathrm{c}} = \frac{S(S+1)}{6}\, A_{\mathrm{f}}\,
  N(E_{\mathrm{F}})\,J_{\mathrm{pd}}^2n_{\mathrm{m}} 
  - k_{\mathrm{B}}T_{\mathrm{AF}} ,
\label{Tc.6}
\ee
where $A_{\mathrm{f}}$ is a Fermi-liquid parameter describing Stoner
enhancement and $T_{\mathrm{AF}}$ is the correction due to superexchange.
The mean-field Curie temperatures of various III-V, II-VI and group-IV
semiconductors doped with a fixed atomic fraction of manganese have been
calculated by Dietl \etal \cite{DOM00,DiO01,DOM01}.
$T_{\mathrm{c}}$ strongly increases for lighter ions for two reasons:
Firstly, the factor $n_{\mathrm{m}}$ increases for smaller ion radii.
Secondly, the effective masses and thus
the density of states typically increase for semiconductors made up of
lighter elements. On the other hand, the exchange interaction
$J_{\mathrm{pd}}$ does not change strongly within each class (III-V, II-VI)
of semiconductors \cite{DOM01}. The trend in $T_{\mathrm{c}}$
qualitatively agrees with experiments. However, in the case of
(Ga,Mn)N, for which a Curie temperature above room temperature is predicted
\cite{DOM00,DiO01,DOM01} and observed \cite{SSS02}, the theory is of limited
validity since manganese forms a deep acceptor level.

Magnetization curves of (Ga,Mn)As have been calculated from this mean-field
theory by K\"onig \etal \cite{KLM00}, Dietl \etal
\cite{DOM01} and Das Sarma \etal \cite{SHK03}, among others. For
large hole concentration a Brillouin-function-like shape of the impurity
magnetization is found. For smaller carrier concentration the curves
cross over to a concave shape over a broad temperature range
\cite{DOM01,SHK03}. On the other hand, the (much smaller) hole contribution
is also Brillouin-function-like for large hole concentration and become
even steeper, with saturation at higher $T/T_{\mathrm{c}}$, for smaller
hole concentrations \cite{SHK03}. Kennett \etal \cite{KBB02}
generalize the VCA/mean-field theory to incorporate a \emph{distribution}
of exchange couplings $J_{\mathrm{pd}}$. They show that a bimodal
distribution of the effective fields experienced by the carriers makes the
concave shape of the impurity-spin magnetization curve even more pronounced
\cite{KBB02}.

While the mean-field theory discussed so far assumes a \emph{static} and
homogeneous effective field, leading to a static and homogeneous self
energy of the carriers, the \emph{dynamical mean-field theory} (DMFT)
\cite{GKK96} drops the assumption of a static field. Thus quantum
fluctuations are partially included. The DMFT has been applied to DMS by
Das Sarma and coworkers \cite{CSM01,HMS02,SHK03}. For small carrier
concentrations $n_{\mathrm{h}}$, the Curie temperature is found to increase
rapidly with $n_{\mathrm{h}}$, similarly to the ordinary mean-field
prediction, but for larger carrier concentration $T_{\mathrm{c}}$ saturates
or falls off again. Also, only for small hole-impurity exchange interaction
$J_{\mathrm{pd}}$ the mean-field result $T_{\mathrm{c}}\propto
J_{\mathrm{pd}}^2$ is found, while $T_{\mathrm{c}}$ increases more slowly
or even decreases for larger $J_{\mathrm{pd}}$ \cite{CSM01}. For the shape
of the magnetization curves, the DMFT results essentially agree with those
of ordinary mean-field theory \cite{SHK03}. It can be criticized that the
Coulomb potential of the acceptors is neglected in \cite{CSM01}. Thus the
quantitative results concerning the formation of an impurity band, its
merging with the valence band and the locatization of holes have to be
interpreted with care.

\subsubsection{Incorporation of disorder}
\label{susus.dis}

The approaches discussed at the beginning of this section neglect disorder.
Disorder enters mainly in two ways: (a) the holes experience the Coulomb
disorder potential of the charged defects and (b) the impurity spins are
localized at the acceptor sites. The inclusion of the Coulomb potential is
complicated by the weak screening. A simple approximation is to consider a
\emph{purely local} potential that depends on the type of impurity
\cite{CGB02,TaK02}. It should be attractive for acceptors (unlike in
\cite{CGB02}) and repulsive for donors. One should be careful in
interpreting the results of such an approach, though. Any property that
crucially depends on the special form of the potential, such as the density
of states of an impurity band, is not easily generalized to real DMS.

Takahashi and Kubo \cite{TaK02} consider (Ga,Mn)As in such an
approximation, where a simple semicircular density of states is assumed for
the valence band and compensating defects are neglected. The distribution
of substitutional manganese is assumed to be random and the dynamical CPA
is employed. For fixed impurity concentration $x=0.05$ the Curie
temperature is found to first increase with hole concentration, as in the
VCA/mean-field approach, but then to decrease again and to vanish for small
compensation $p=p_{\mathrm{c}}\lesssim 1$ \cite{TaK02}. The decrease of
$T_{\mathrm{c}}$ is an effect of the hole density of states, which for
weaker doping can be understood from the properties of the impurity band:
The width of the impurity band is found to increase with impurity spin
polarization. Thus for small hole concentrations the hole energy is reduced
in the ferromagnetic state, whereas for large concentrations the reduction
is smaller and eventually becomes an increase, destroying ferromagnetism
\cite{TaK02}. The authors \cite{TaK02} suggest that the same picture also
applies for higher doping ($x=0.05$). This mechanism is described as
\emph{double exchange} but not, as usual, in a narrow d-band but in a
narrow impurity band \cite{TaK02}. It should be checked how the results
change for an impurity potential of longer range and with compensating
defects included.

Disorder due to a weakly screened Coulomb potential and the distribution
of impurity spins is studied in \cite{TSO,icps,pasps}.
The Hamiltonian of holes in the Coulomb disorder
potential is augmented by the exchange interaction with the impurity spins
$\bi{S}_i$ ($S=5/2$),
\be
H = \sum_{n\sigma} \xi_n\, c_{n\sigma}^\dagger c_{n\sigma}
  - J_{\mathrm{pd}} \sum_i \bi{s}_i \cdot \bi{S}_i ,
\ee
where
\be
\bi{s}_i \equiv \sum_{n\sigma n'\sigma'} c_{n\sigma}^\dagger
  \psi_n^\ast(\bi{R}_i)\,
  \frac{\mbox{\boldmath$\sigma$}_{\sigma\sigma'}}{2}\,
  \psi_{n'}(\bi{R}_i) c_{n'\sigma'}
\end{equation}
are hole spin polarizations at the manganese sites \cite{icps}.
$\xi_n$ and $\psi_n$ are
the hole eigenenergies and eigenfunctions in the absence of exchange but
including the Coulomb disorder potential.
The Hamiltonian is decoupled in a mean-field approximation, where,
importantly, no spatial average is taken.
Details of the formal derivation within
the functional-integral approach are given in \cite{icps}. The mean-field
decoupling is very similar to the one employed in \cite{BB}, which is,
however, concerned with the light-doping regime. In \cite{TSO,icps,pasps} a
\emph{collinear} magnetization is assumed.
One can show that the equations for $T_{\mathrm{c}}$, obtained by linearizing
the mean-field equations, are unchanged by dropping this assumption.

\begin{figure}
\centerline{\includegraphics[width=7cm,clip]{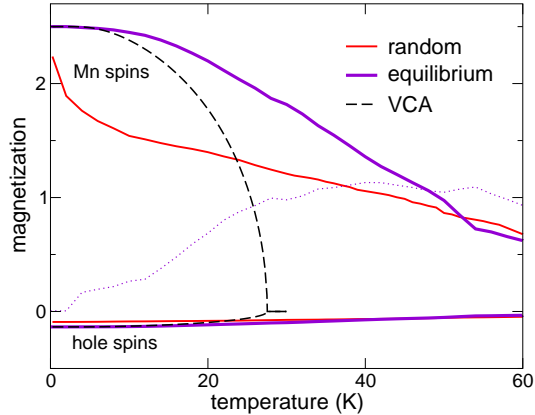}}
\caption{\label{fig.magn}Magnetization of manganese
(up) and hole spins (down) as functions of temperature for
$x=0.05$ and $p=0.3$ for random and equilibrated (clustered)
configurations, after \protect\cite{TSO,icps,pasps}.
The dotted curve gives the standard deviations of the manganese spin
polarizations. For comparison, the long-dashed curves show the
magnetizations obtained from a theory that totally neglects disorder.}
\end{figure}

Figure \ref{fig.magn} shows the magnetization curves for manganese and hole
spins for $x=0.05$ and $p=0.3$ for random and equilibrium defect
configurations \cite{TSO,icps,pasps}. Since these results are obtained from
a mean-field theory for a disordered ferromagnet, one should not trust them
at temperatures $T\sim T_{\mathrm{c}}$. The shape of the impurity-spin
magnetization curve is anomalous for random defects, showing a rapid
initial decay and then a long linear region. The shape becomes convex and
more Brillouin-function-like with annealing \cite{TSO}, except for a tail
at higher temperatures. This qualitatively agrees with experimental
annealing studies \cite{Potashnik}. The anomalous shape for random defects
is due to the localization tendency of the holes, which leads to a
shorter-range effective manganese-manganese spin interaction and thus to a
broad distribution of effective fields acting on these spins. For higher
temperatures, only a few spins with strong interactions carry most of the
magnetization, leading to a long tail for random defects. \Fref{fig.magn}
also shows the standard deviation of the manganese spin polarizations. The
standard deviation becomes comparable to the average manganese spin
polarization at intermediate temperatures. This shows that the decrease of
the magnetization with increasing temperature is initially dominated by
disordering of large moments and not by their reduction \cite{icps}. The
hole-spin magnetization curve has a more normal shape. Note that the holes
are not completely spin polarized at $T\to 0$ since the Fermi energy is
larger than the effective Zeeman energy.

\Fref{fig.magn} also compares the results with the disorder-free theory
\cite{DHM97,JAL99,DOM00}. Note that the Curie temperatures are
significantly \emph{enhanced} by disorder. It is an important question
whether this is an artefact of mean-field theory or a real physical effect.
For random defect positions and correspondingly shorter-range
manganese-manganese interactions, the mean-field $T_{\mathrm{c}}$ is
governed by atypically \emph{strong} couplings. (In the extreme limit,
mean-field theory incorrectly predicts a nonzero $T_{\mathrm{c}}$ if only a
single coupling is nonzero.) Thermal fluctuations destroy the long-range
order in this case, reducing $T_{\mathrm{c}}$. On the other hand, for
clustered defects the hole states are extended, the effective
manganese-manganese interactions are of longer range and mean-field theory
is much more appropriate. Thus in the metallic regime the enhancement of
$T_{\mathrm{c}}$ might be real.

It is also interesting that in
$\mathrm{UCu}_2\mathrm{Si}_{2-x}\mathrm{Ge}_x$, where \emph{electronic}
disorder is controlled by $x$ whereas the magnetic uranium ions always
form a regular lattice, $T_{\mathrm{c}}$ is enhanced by electronic disorder
\cite{SiC03,SCM03}. Due to the coupling between impurity spins and
carriers, disorder here leads to the damping of spin fluctuations which
reduces their effect on $T_{\mathrm{c}}$ \cite{SiC03}. The same mechanism
might also apply to the more complicated case of DMS \cite{SiC03,SCM03}.

Mean-field magnetization curves obtained for various manganese
concentrations and fully equilibrated defects \cite{TSO,icps} show a
crossover from a Brillouin-function-like shape at $x=0.05$ to a
concave shape at $x=0.01$, as expected for insulating DMS.

\begin{figure}[ht]
\centerline{\includegraphics[width=7.5cm]{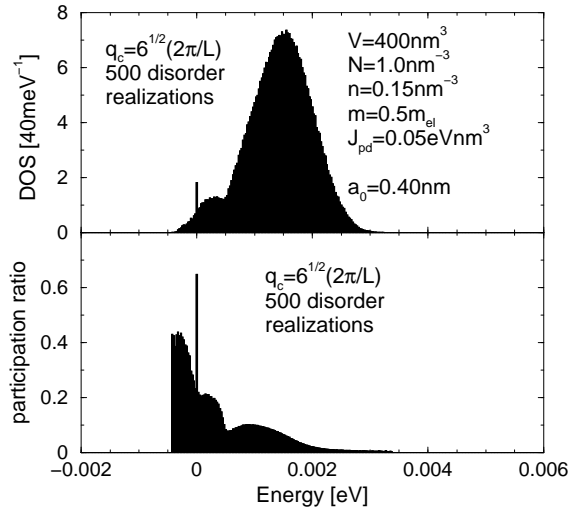}}
\caption{\label{fig.mDOS.JS}Density of states of collective
magnetic excitations averaged over disorder realizations (upper panel) and
averaged participation ratio of these states (lower panel). The
participation ratio of the magnetic excitations can be interpreted as the
fraction of spins that are actually taking part in an excitation. The sharp
peak at zero energy corresponds to the Goldstone mode of uniform rotation of
all spins. From \protect\cite{ScM02}.}
\end{figure}

Schliemann and MacDonald \cite{ScM02,Sch03} study the stability of the
collinear mean-field magnetization. To that end they go beyond mean-field
theory by deriving the spin-fluctuation propagator and from this the
energies of elementary magnetic excitations, similarly to K\"onig \etal
\cite{KLM00,KLM01,KSJ03}. However, in \cite{ScM02,Sch03} the disorder in
the impurity positions is retained. The positions are assumed to be random.
For a parabolic valence band it is shown that the collinear mean-field
solution is stationary but that a fraction of the excitations
generically appear at \emph{negative} energies, as shown in
\fref{fig.mDOS.JS} \cite{ScM02}. This means that the collinear saddle point
is not stable. \Fref{fig.mDOS.JS} also shows that the modes with negative
energy involve many impurity spins.

\begin{figure}[ht]
\centerline{\includegraphics[width=7cm,clip]{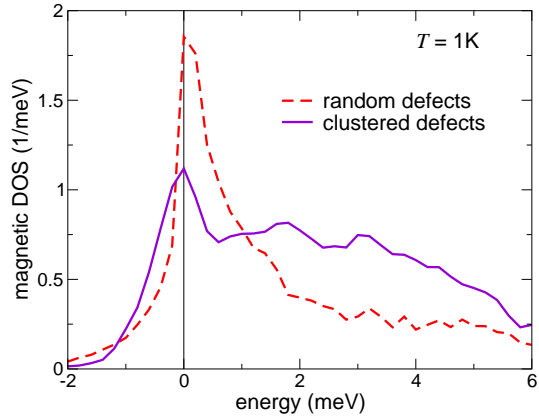}}
\caption{\label{fig.mDOS2}Density of states of magnetic excitations,
assuming a collinear mean-field solution, calculated from the model of
\protect\cite{TSO}. The dashed (solid) line corresponds to the case of
random (clustered) defects. The excitations are given a finite width of
$0.1\,\mathrm{meV}$ to obtain smooth curves.}
\end{figure}

Similarly to the model of \cite{AMD02,AlD03}, the Coulomb disorder
potential is neglected in \cite{ScM02,Sch03} so that the localization
properties of the holes are probably not correctly described. To take them
into account, we can start from the model of \cite{TSO}, derive the
mean-field approximation more formally using a Hubbard-Stratonovich
decoupling and include Gaussian fluctuations in the decoupling fields. The
details will be presented elsewhere \cite{elsewhere}. The main result is
apparent from \fref{fig.mDOS2}: The excitations at negative energies
persist if the Coulomb disorder potential is included, for both random and
clustered impurities. It is also seen that the equilibration of defects and
the corresponding reduction of disorder leads to a shift of weight to
higher energies. This means that the magnetic system becomes \emph{stiffer}
with annealing, which is consistent with longer-range interactions. This
stiffening has also been found by Singh \cite{Sin03}.

While in \cite{ScM02} and in \fref{fig.mDOS2} a parabolic band is
considered, the stability of the collinear state in the Kohn-Luttinger
6-band model is studied in \cite{Sch03}. Whereas the collinear state is
stationary but unstable for a parabolic band, it is not even stationary for
the 6-band model \cite{Sch03}, \textit{i.e.}, in a collinear configuration
a transverse \emph{force} acts on the impurity spins. In particular, it is
not the ground state. This is due to the interplay of spin-orbit coupling
and disorder \cite{Sch03}. The deviation of the true ground state from
collinearity may be small, though \cite{John.priv}.

Another study of the effect of disorder has been presented by Chudnovskiy
and Pfannkuche \cite{ChP02}. The model consists of holes on a lattice and
impurity spins. A random number, taken to be five on average, of impurity
spins is associated with every lattice site and assumed to have an exchange
interaction with a hole at that site. Using a functional-integral approach
and averaging over disorder with the replica trick, mean-field
magnetization curves and estimates for $T_{\mathrm{c}}$ are calculated for
various values of the \emph{variance} of the number of impurity spins
associated with each site \cite{ChP02}. Increased variance, \textit{i.e.},
stronger disorder, is found to lead to a higher Curie temperature
\cite{ChP02}. However, the model is not easy to relate to real DMS.

\subsubsection{Beyond mean-field theory}

It is important to check the applicability of mean-field theory. Schliemann
\etal \cite{SKM01,SKL01} perform hybrid MC simulations for holes
interacting with classical impurity spins. They mostly consider a
parabolic-band model but also a 6-band Kohn-Luttinger Hamiltonian. The
impurity spins are placed at fully random positions without regard of the
lattice. The Coulomb disorder potential is neglected. A Gaussian form of
the spatial dependence of the hole-impurity exchange interaction is assumed
phenomenologically. We comment on this assumption in \sref{sus.spinonly}.
Unlike in \cite{AMD02,AlD03}, the numerical effort due to the
diagonalization of the carrier Hamiltonian is reduced by using a hybrid MC
method, where all spins are updated at every step \cite{SKM01}. Up to 540
impurity spins are included. The Curie temperature is strongly reduced by
fluctuations \cite{SKM01,SKL01}. Above $T_{\mathrm{c}}$ a phase with large
local moments but no long-range order is found \cite{SKM01}. The
magnetization curves for both impurity and carrier spins are found to be
rather normal (Brillouin-function-like) for the parabolic-band case and
slightly less so, with a nearly straight region, for the 6-band model. The
magnetization is generally not collinear for $T\to 0$, although this effect
is rather small, especially for the 6-band model \cite{John.priv}. For the
parabolic band, the curves look very different from the mean-field results
discussed above.

Alvarez \etal \cite{AMD02,AlD03} also perform MC simulations for holes and
classical impurity spins. The model has been discussed in
\sref{sus.trans.band}. The resulting mag\-ne\-ti\-za\-tion curves are quite
anomalous for all parameters, nearly straight or concave in the whole
temperature range up to $T_{\mathrm{c}}$, see, \textit{e.g.}, \fref{fig.AD}
(upper panel) \cite{AMD02}. The origin of the discrepancy compared to
Schliemann's work \cite{SKM01} is not clear. The tight-binding instead of
parabolic band should not change the results qualitatively for the small
Fermi seas considered here. The main technical differences seem to be (a)
that \cite{SKM01} allows the impurity spins to come arbitrarily close,
whereas \cite{AMD02} contains a natural lattice cutoff for the separation,
and (b) that \cite{SKM01} treats much larger system sizes (up to 540
impurities compared to up to about 50 in \cite{AMD02}). Whether the
impurities sit on correct cation positions or are randomly distributed in
the continuum does not seem to make an appreciable difference, though
\cite{Sch03}. Further MC studies are definitely necessary, also to study
the effect of clustered defects.

The dependence of $T_{\mathrm{c}}$ on $J_{\mathrm{pd}}$ is also discussed
in \cite{AMD02}. For large $J_{\mathrm{pd}}$ the Curie temperature
decreases since the exchange interaction becomes strong enough to localize
the holes \cite{AMD02,AlD03}, as discussed in \sref{sus.trans.band}. If
holes at the Fermi energy become localized a description in terms of
percolation should become applicable.

\subsection{Percolation picture}

The application of the percolation picture to
DMS \cite{BB,BBlong,KaS02,MAD02,AMD02,SHK03,Erwin,KS,FZD03}
extends earlier results for dilute ferromagnetic \emph{metallic} alloys
\cite{KSS73}. As discussed in \sref{sus.crossover}, the BMP's overlap for
sufficiently high impurity and hole concentrations. This leads to an
effective ferromagnetic interaction between impurity spins that are part of
the same system of overlapping BMP's. At $T\to 0$ we
expect a percolation transition as a function of impurity and hole
concentrations, at which an infinite system of overlapping BMP's appears.
Two BMP's are here said to overlap if the effective ferromagnetic
interactions between the localized spins is so strong that their
correlation is not destroyed by quantum fluctuations. To
the best knowledge of the author, this quantum phase transition has not
been studied to date. At finite temperatures, the ferromagnetic transition
in this regime has been considered by Kaminski and Das Sarma
\cite{KaS02,KS}. The main idea of this work is the following: The effective
ferromagnetic interactions are inhomogeneous even for randomly distributed
impurities. Therefore at a given temperature $T>T_{\mathrm{c}}$ only those
impurity spins will align ferromagnetically that are coupled by effective
interactions that thermal fluctuations do not overcome. In \cite{KaS02,KS}
this physics is expressed by a temperature-dependent radius of BMP's. As
the temperature is lowered, the size of the ferromagnetically aligned
regions grows and eventually becomes infinite at $T_{\mathrm{c}}$. The
Curie temperature obtained by this percolation approach is \cite{KaS02,KS}
\be
T_{\mathrm{c}} \approx a_{\mathrm{B}} \sqrt{n_{\mathrm{m}}}\,
n_{\mathrm{h}}^{-1/6}\, sS\, |J_{\mathrm{pd}}|\,
\exp\!\left(-\frac{0.86}{a_{\mathrm{B}} n_{\mathrm{h}}^{1/3}}\right) ,
\ee
where $s$ and $S$ are the hole and impurity
spin quantum numbers, respectively. For small hole concentration $n_{\mathrm{h}}$ the Curie
temperature becomes exponentially small. Since quantum fluctuations are not
included, $T_{\mathrm{c}}$ goes to zero only for $n_{\mathrm{h}}\to 0$.

Essentially the same physics is discussed by Alvarez \etal \cite{AMD02}.
They denote the temperature where ferromagnetic domains form by $T^\ast$.
For $T_{\mathrm{c}}<T<T^\ast$ the magnetizations of the domains do not have
long-range correlations. This is called a \emph{clustered state} by Alvarez
and Dagotto \cite{AlD03a}, who discuss this state as a special case
of a more general paradigm that also applies to manganites and cuprates.

In the strongly localized regime the magnetization as a function of
temperature has been obtained numerically using a mean-field approximation
\cite{BB} and MC simulations \cite{MAD02} and analytically within a
percolation theory \cite{KaS02} and a mean-field approximation for
localized holes \cite{SHK03}. In the numerical mean-field calculation the
exchange interaction between hole impurity spins is decoupled at the
mean-field level, while the disorder due to the (fully random) distribution
of magnetic impurities is retained \cite{BB,BBlong}. This approach is
limited to rather low doping \cite{comment,reply}. In \cite{BB,BBlong} the
sign of the hopping integral \eref{TofR.2} had to be inverted by hand to
obtain reasonable results for the magnetization \cite{comment,reply}.
Correcting the sign of $T(\bi{R})$ inverts the \emph{highly assymmetric}
impurity band, which drastically reduces the density of states at the Fermi
energy, and, consequently, the mean-field Curie temperature \cite{comment}.
This again indicates that a model with hydrogenic impurity states is not
really sufficient.

The impurity magnetization curve of \cite{BB,BBlong} is concave over a
broad tem\-pe\-ra\-ture range, whereas the hole magnetization saturates
quickly below $T_{\mathrm{c}}$. The mean-field Curie temperature is found
to be enhanced by disorder by a factor of about two relative to a periodic
superlattice of manganese ions \cite{BB,BBlong}. These results are similar
to the metallic regime discussed above. Since the effective interaction
between impurity spins is of very short range here, the remark made there
carries over: Mean-field theory tends to overestimate $T_{\mathrm{c}}$
since it overemphasizes anomalously strong couplings. Thus the enhancement
of $T_{\mathrm{c}}$ by disorder seems questionable in the localized regime
\cite{LJM02}. In fact MC simulations performed for the same model
\cite{MAD02} do not find a significant enhancement of $T_{\mathrm{c}}$ by
disorder. The MC results for the magnetization curves also show an extended
concave region but differ from the mean-field results in that they appear
to stay concave up to $T_{\mathrm{c}}$ \cite{MAD02}, similarly to MC
simulations in the metallic regime \cite{AMD02}.

In the MC simulations \cite{MAD02} the impurity magnetization for $T\to 0$
is much smaller than the saturated value. The origin discussed in
\cite{MAD02} is that isolated impurity spins and small clusters are
practically unoccupied by holes and do not order for any $T>0$. However,
even these impurity spins should experience an effective Zeeman field from
the holes; the probability of finding a hole there is exponentially small
but not zero. Thus for sufficiently small temperatures these impurity spins
would come into alignment with the rest, but these
temperatures might be unobservable in simulations and in experiments. Perhaps
more relevant for real DMS, quantum fluctuations---not included in the MC
simulations---may destroy this order even at $T\to 0$.

The shape of both the impurity and hole magnetization curves from \cite{BB}
is well reproduced by analytical results of the percolation \cite{KaS02}
and mean-field \cite{SHK03} approach, indicating that the qualitative
features of the magnetization are rather robust and do not depend on the
details of the model \cite{comment}. Interestingly,
the percolation theory \cite{KaS02} gives a \emph{universal} expression for
the magnetization, determined by the volume of the infinite ferromagnetic
domain. The concave shape of the impurity magnetization curve found in
mean-field theories \cite{BB,SHK03} is probably not an artefact since MC 
simulations \cite{MAD02} and percolation theory \cite{KaS02} find a similar
shape, even though the actual value of $T_{\mathrm{c}}$ is strongly
overestimated by mean-field theory \cite{KLM00,YaM03}. The concave shape
becomes less pronounced with increasing concentration of impurity spins
\cite{KaS02}. These results are qualitatively similar to the ones obtained
from the band picture, see \sref{sus.mag.band}, and also to the case of
dilute ferromagnetic metals \cite{KSS73}.

The Curie temperature $T_{\mathrm{c}}$ is predicted to decrease for large
exchange interactions $J_{\mathrm{pd}}$ \cite{AMD02}. One origin of this
effect is that the holes become more strongly bound to the impurities due
to this exchange interaction. Thus the overlap between BMP's is reduced,
their effective exchange coupling is weakened and $T_{\mathrm{c}}$
decreases. As noted above, the holes are eventually localized if
$J_{\mathrm{pd}}$ becomes sufficiently large, but this is not the dominant
mechanism in DMS.

For \emph{clustered} magnetic defects the percolation picture
\cite{KS,AMD02} naturally predicts the Curie temperature to be dominated by
the weak effective coupling \emph{between} clusters. The mean-field
approximation \cite{BB,TSO,ScM02,Sch03} becomes increasingly inaccurate for
pronounced clustering if the effective interaction between impurity spins
is of short range. The reason is that in mean-field theory $T_{\mathrm{c}}$
is governed by anomalously strong couplings within the clusters, whereas
the true $T_{\mathrm{c}}$ is determined by the \emph{weak} coupling between
clusters. The condition of short-range interactions is crucial, see the
discussion for the metallic case above. Quantitative results from
percolation theory for $T_{\mathrm{c}}$ or the temperature dependence of
the magnetization in the clustered case do not seem to exist. However, one
can expect that large ferromagnetically ordered domains exist at $T\gtrsim
T_{\mathrm{c}}$, which align at $T_{\mathrm{c}}$ \cite{KS}. This should
lead to a more rapid increase of the magnetization below $T_{\mathrm{c}}$
than for uncorrelated impurities. The same qualitative result is found for
clustered defects in the metallic regime (but for a different reason---the
disorder potential is reduced by clustering), \textit{cf}.\
\sref{susus.dis}.

For $T\gtrsim T_{\mathrm{c}}$ the domains are easily aligned by an applied
magnetic field. In this situation the weak inter-cluster coupling becomes
irrelevant and the magnetization is dominated by the strong coupling within
the clusters. This should lead to a Brillouin-function-like magnetization
curve, as observed in some experiments \cite{EWC02,SSS02,BlW02,BlW03}.

One important conclusion is that the crossover from a
``normal'' magnetization curve to a concave one for decreasing carrier
concentration or increasing localization is very robust. This agrees with
the experimental observation that the magnetization curves change smoothly
with growth parameters and annealing time \cite{Potashnik,MSS02},
and thus with disorder strength. The experimental and theoretical results
indicate that the $T=0$ metal-insulator transition is not accompanied by a
magnetic transition. Conversely, due to this robustness comparison of
calculated magnetization curves with experiment is not sufficient to
decide between various models or approximations.

\subsection{Spin-only models}
\label{sus.spinonly}

We now turn to disorder effects in \emph{spin-only} models of
DMS. It is an attractive proposition to remove the carriers from the model
and incorporate their effect into the interaction between impurity spins.
In principle, this can be done by integrating out the carriers,
\textit{e.g.}, in the functional-integral formalism
\cite{KLM00,KLM01,KSJ03}. Employing a parabolic-band model without disorder
and neglecting three-spin and higher interactions, one obtains the standard
Ruderman-Kittel-Kasuya-Yosida (RKKY)
interaction \cite{Yosida}, as shown, \textit{e.g.}, by Dietl
\etal \cite{DHM97}. The effective interaction $J_{ij}$ between
spins $\bi{S}_i$ and $\bi{S}_j$ is an oscillating function of their
separation, which leads to frustration. The typical period of these
oscillations is determined by the Fermi wave length $\lambda_{\mathrm{F}}$,
which in DMS is typically \emph{larger} than the distance between impurity
spins.
Thus their interaction is dominated by the first, ferromagnetic section of
the oscillations, making ferromagnetic long-range order possible.
In the limit that the typical separation is small compared to
$\lambda_{\mathrm{F}}$, this leads to essentially the same $T_{\mathrm{c}}$
as given in equation \eref{Tc.4} \cite{DHM97}.
For a random distribution of defects,
the antiferromagnetic interactions dominate for \emph{some} of them. These
align in \emph{antiparallel} to the total magnetization \cite{GPS85}, which
leads to a reduced magnetization for $T\to 0$. For clustered spins this
effect probably becomes even stronger, since now the typical separations
between clusters become important, which are larger than the typical
separations between random impurities. For weak compensation,
\textit{i.e.}, large $p\approx 1$, more and more spins are
antiferromagnetically coupled, eventually leading to spin-glass behaviour
instead of ferromagnetism. In addition, as K\"onig \etal
\cite{KLM00,KLM01,KSJ03} point out, the Zeeman energy in the effective
field is smaller than the Fermi energy but not necessarily negligible so
that one should take the hole spin polarization into account in calculating
the RKKY interaction.

It is interesting that K\c{e}pa \etal \cite{KKB03} found in
inelastic neutron scattering experiments on the II-VI material (Ze,Mn)Te
that only a model of \emph{itinerant} carriers \cite{DHM97} is in rough
agreement with the measured carrier contribution to the RKKY-like
interaction. A BMP model of strongly localized holes does not fit the data.
This is the case although the samples are insulating at low temperatures
\cite{KKB03}. Presumably the localization length is larger than the typical
range of the RKKY interaction.

Zar\'and and Jank\'o \cite{ZaJ02} employ the Kohn-Luttinger Hamiltonian
\cite{KL} for GaAs restricted to heavy and light holes in the spherical
approximation. Integrating out the holes, they obtain the effective
interaction between manganese spins \cite{ZaJ02}. The hole polarization by
other impurities is neglected here, as is the Coulomb disorder potential.
Thus no disorder is present in the hole Green function.
The interaction shows the
usual oscillations with distance and, more interestingly, a pronounced
anisotropy: In the effective interaction
\be
H_{\mathrm{eff}} = -K_\|(|\bi{R}_1-\bi{R}_2|)\, S_1^\| S_2^\|
  - K_\perp(|\bi{R}_1-\bi{R}_2|)\, \bi{S}_1^\perp \cdot \bi{S}_2^\perp ,
\ee
where $S_i^\|$ and $\bi{S}_i^\perp$ are the components of the impurity spin
$\bi{S}_i$ parallel and perpendicular to $\bi{R}_1-\bi{R}_2$, respectively,
the exchange parameters $K_\|$ and $K_\perp$ are typically very different,
often even in their sign \cite{ZaJ02}. This leads to frustration even in an
ordered arrangement of impurity spins. The origin of the anisotropy is the
spin-orbit coupling in the valence band; without spin-orbit coupling there
would not be any coupling between directions in spin space ($\bi{S}_i$) and
real space ($\bi{R}_1-\bi{R}_2$).

In the second step, MC simulations are performed for a classical spin-only
model containing the anisotropic effective interaction \cite{ZaJ02}. The
manganese impurities are randomly distributed on the face-centered-cubic
cation lattice. For $x=0.05$ the impurity magnetization curve is nearly
straight for lower temperatures and becomes concave close to
$T_{\mathrm{c}}$, in qualitative agreement with mean-field results and the
MC simulations of Alvarez \etal \cite{AMD02,AlD03}. The impurity spins are
not collinear for $T\to 0$. On the contrary, the angle between individual
spins and the average magnetization direction shows a broad distribution,
in particular for large hole fraction $p$, as a consequence of disorder
together with the anisotropic interaction.

These results have been criticized by Brey and G\'omez-Santos \cite{BrG03},
who find that the spin anisotropy of the effective impurity-spin
interaction is always smaller than 5\%. They employ a 6-band Kohn-Luttinger
Hamiltonian for GaAs without additional approximations. In particular, this
means that the Fermi surface is far from spherical \cite{DOM00,DOM01},
unlike in \cite{ZaJ02}. The second difference is that in \cite{BrG03} a
phenomenological Gaussian form of the hole-impurity exchange interaction
$J_{\mathrm{pd}}$ is assumed, whereas in \cite{ZaJ02} this interaction is
purely local. The width of the Gaussian is chosen of the order of the
nearest-neighbour separation, but the results are found to depend
significantly on this width \cite{BrG03}. Due to the $\bi{k}$-space cutoff
imposed by this width, the RKKY-like interaction is found to be nearly
isotropic \cite{BrG03}, despite the highly non-spherical Fermi surface. The
validity of the Gaussian $J_{\mathrm{pd}}$ has to be checked.

Motivated by the apparent smallness of the anisotropy, a disordered
Heisenberg model with isotropic (in spin space and real space) interactions
is studied in \cite{BrG03}. The magnetization curves obtained by MC
simulation are quite Brillouin-function-like, except at low temperatures,
where they approach full polarization with a finite slope \cite{BrG03}.
Because of the isotropic interaction, the magnetization is collinear for
$T\to 0$. The results agree quite well with the MC simulations of
Schliemann \etal \cite{SKM01} for essentially the same model, including the
Gaussian form of $J_{\mathrm{pd}}$. However, in \cite{SKM01} the carrier
Hamiltonian is diagonalized at every MC step, whereas in \cite{BrG03} the
holes are integrated out first. The agreement supports the treatment in
\cite{BrG03}, but does of course not say anything about the applicability
of the model. The magnetization curves do not agree with the anomalous
results discussed above, indicating that the model underestimates the
effect of disorder.

The magnetic collective excitations are also considered in \cite{BrG03}.
Their density of states is obtained and averaged over impurity
realizations. For the effective spin interaction calculated for a
parabolic-band model, it shows excitations at negative energies, consistent
with \cite{ScM02}, see \fref{fig.mDOS.JS}. These are shifted to positive
energies for a broader Gaussian $J_{\mathrm{pd}}$ \cite{BrG03}, since this
Gaussian damps the RKKY-like oscillations and thus reduces frustration. For
the 6-band Kohn-Luttinger Hamiltonian no excitations at negative energies
appear \cite{BrG03}, due to the finite width of $J_{\mathrm{pd}}$ together
with the non-spherical Fermi surface. In those cases the collinear ground
state is thus found to be stable. However, Schliemann \cite{Sch03}
has shown that the collinear state is not stationary for the 6-band model
and thus cannot be the ground state. Taken together, these results suggest
that the ground state weakly deviates from collinearity.

A problem of all spin-only models discussed so far is the neglection of
disorder in the calculation of the RKKY-like effective spin interaction.
Priour \etal \cite{PHS03} assume a free-carrier RKKY interaction between
impurity spins \cite{Yosida} but include the effect of electronic disorder
and a finite mean free path by a phenomenological exponential damping
factor. They also add an antiferromagnetic nearest-neighbour interaction
and assume a random distribution of impurity spins. A mean-field
approximation that keeps the spatial disorder \cite{BB,TSO} is then used to
obtain magnetization curves and $T_{\mathrm{c}}$ \cite{PHS03}. They find a
crossover from convex to concave magnetization curves with decreasing
cutoff length in the RKKY interaction \cite{PHS03}. The Curie temperature
is strongly reduced relative to the VCA result \cite{DHM97,JAL99,DOM00} by
a small cutoff length (strong disorder).

Zhou \etal \cite{ZKW03} consider classical disordered Heisenberg models
with various functional forms of the interaction. The effect of anisotropy
in spin space and of electronic disorder, which is described by the width
of the distribution of coupling constants, is studied with extensive MC
simulations, motivated by the results of Zar\'and and Jank\'o \cite{ZaJ02}.
Random distributions of impurity spins on the fcc cation sublattice are
assumed. Anisotropy is found to be irrelevant for the magnetization at high
temperatures and to lead only to a small reduction at $T=0$, in contrast to
\cite{ZaJ02}, regardless of the form of the interaction (short range vs.\
long range, RKKY-like vs.\ purely ferromagnetic) \cite{ZKW03}. On the other
hand, the frustration due to an oscillating RKKY interaction leads to a
large reduction of the magnetization at $T=0$. The effect of electronic
disorder is found to be rather weak, but the magnetization curves become
less Brillouin-function-like and more linear for large disorder
\cite{ZKW03}. It should be kept in mind that electronic disorder is
incorporated phenomenologically by a distribution of coupling constants,
which is not claimed to be realistic in \cite{ZKW03}.

The next step is the incorporations of disorder at the CPA level. Bouzerar
\etal \cite{BKB02} employ the CPA for a one-band tight-binding model
without Coulomb disorder potential to evaluate the RKKY-like interaction,
taking into account that the impurity spins are located at manganese sites.
The hole-impurity exchange interaction $J_{\mathrm{pd}}$ is assumed to be
local. For increasing $|J_{\mathrm{pd}}|$ the Curie temperature overshoots
the disorder-free value and then rapidly drops to zero \cite{BKB02}. Since
the CPA does not include localization, the origin cannot be localization of
holes, unlike in \cite{AMD02,AlD03}. Rather, for large $|J_{\mathrm{pd}}|$
the impurity band split off the valence band and the density of states at
the Fermi energy vanishes. It is also shown that the sign of
$J_{\mathrm{pd}}$ is important for $T_{\mathrm{c}}$, while the
disorder-free RKKY theory just gives $T_{\mathrm{c}}\propto
J_{\mathrm{pd}}^2$ \cite{BKB02}. In \cite{BKB03}, the RKKY-like interaction
is evaluated for a more realistic band structure of (Ga,Mn)As obtained from
an \textit{ab-initio} calculation with a CPA-treatment of disorder
(substitutional manganese and antisites). The Coulomb disorder potential is
neglected. The resulting impurity-spin interaction remains ferromagnetic up
to large separations, in contrast to, \textit{e.g.}, \cite{ZaJ02}. This is
probably why fluctuations are not found to suppress $T_{\mathrm{c}}$
dramatically. The partly contradictory results on the RKKY-like interaction
in metallic DMS show that further work is necessary.

Erwin and Petukhov \cite{Erwin} derive a spin-only model in the opposite
\emph{weak-doping} limit. The effective impurity-spin interaction is
calculated in second order in the hopping amplitude between hydrogenic
impurity states. It falls off exponentially since the hopping amplitude
does, see equation \eref{TofR.2}. Standard percolation theory \cite{Ios95}
is then used to obtain an approximate Curie temperature. Interestingly,
$T_{\mathrm{c}}$ vanishes for zero compensation, \textit{i.e.}, one hole
per impurity, since for ferromagnetically aligned hole spins the holes then
cannot hop. However, it should be kept in mind that this approach is
limited to the dilute regime.

A quite different approach towards a simpler effective model is to
integrate out the \emph{impurity spins}, not the carriers. This is done by
Santos and Nolting \cite{SaN02} for a disorder-free Kondo-lattice model and
by Galitski \etal \cite{GKS03} for DMS in the strongly localized regime. In
the limit of strong compensation, the system consists of nearly isolated
BMP's made up of a single carrier and several impurity spins, see
\sref{sus.crossover}. Starting from the percolation picture \cite{KaS02}
the impurity spins can be integrated out, leading to an effective
Heisenberg model for BMP's \cite{GKS03}. The effective exchange interaction
is always ferromagnetic but depends exponentionally on separation, making
the Heisenberg model strongly disordered. A random distribution of magnetic
impurities is assumed \cite{GKS03}. It is then proved that the paramagnetic
phase is a \emph{Griffiths phase} \cite{Gri69,McW68} in this model,
\textit{i.e.}, the magnetization is a (weakly) non-analytic function of the
external magnetic field due to rare extended magnetic domains. Much
stronger Griffiths-McCoy singularities \cite{Bra88} are derived for the
local dynamic susceptibility in the paramagnetic phase \cite{GKS03}.
Indeed, Galitski \etal \cite{GKS03} suggest that insulating DMS may be an
ideal system to study Griffiths effects. It would also be interesting to
study the Griffiths-McCoy singularities as the DMS is tuned towards the
metallic regime.

\section{Conclusions}
\label{sec.conc}

By now it has been widely recognized that disorder plays an important role
in DMS, as the works discussed in this paper attest. The detailed
distribution of defects (random vs.\ clustered) is crucial, since it
strongly affects both the Coulomb disorder potential and the additional
disorder due to the positions of impurity spins. Some approaches have
neglected disorder or included only the disorder due to the distribution of
impurity spins. But for a realistic description of transport the inclusion
of the dominant source of scattering, \textit{i.e.}, the Coulomb disorder
potential, is crucial, even in the metallic phase. The same holds for the
magnetic properties since the magnetism is carrier-mediated, as is clearly
shown by experiments.

A multitude of models have been proposed and are treated by a variety of
methods. Practically all theoretical studies start from either a band model
or from a model of isolated impurity states. It is important to note,
however, that the band model can in principle describe not only the
metallic but also the insulating regime if disorder is properly taken into
account. It is thus the more general description.

The shape of the magnetization curve has emerged as a standard yardstick
for theories of DMS. However, the concave shape of the impurity
magnetization curve is not \emph{only} an effect of disorder, since it
appears also for disorder-free models. It tends to be more pronounced if
disorder is included, though. In this case the magnetization curves become
more concave for less metallic or more insulating DMS, regardless of the
model and the approximations used. On the other hand, the strong
enhancement of $T_{\mathrm{c}}$ by disorder predicted by mean-field
theories for both metallic and insulating DMS is \emph{not} seen in MC
studies. This indicates that it is really an artefact of mean-field theory,
although the issue is not yet fully settled. Transport measurements are, in
principle, better suited to learn about disorder in DMS. The problem here
is that theory is at present lacking behind experiment. For example, no
generally accepted theory for the resistivity maximum around
$T_{\mathrm{c}}$ exists.

We conclude with listing directions of research that could advance the
theory of DMS at the present stage: Firstly, the nature of the impurity
states in DMS with typical impurity concentrations of a few percent should
be analyzed in detail. Are they really hydrogenic? The answer is crucial
for building valid models.

More detailed simulations of DMS growth, starting from a microscopic model
and perhaps employing \textit{ab-initio} calculations of configuration
energies, would be very useful to understand the formation of the various
defect species during low-temperature MBE. Further improvements in
\textit{ab-initio} methods are required with the goal to obtain shallow,
\emph{hydrogenic} impurity states in the weak-doping regime. One can then
hope to describe the crossover to the heavy-doping limit correctly and to
reproduce the qualitative difference between, say, (Ga,Mn)As and (Ga,Mn)N.

Also, a CPA treatment of DMS with a proper screened Coulomb potential would
be desirable. Up to now all CPA treatments suffer from the assumption of a
purely local Coulomb potential.

The inclusion of \emph{quantum} fluctuations would also be very useful, for
example in quantum MC simulations. As discussed above, quantum fluctuations
are expected to destroy any ferromagnetic order at low impurity
concentration. Simulations would provide information about the
corresponding quantum phase transition, presumably of percolation type, as
a function of concentrations and strength of disorder.

Finally, a transport theory for DMS is still missing. In particular, there
is no agreement on the origin of the resistivity maximum close to
$T_{\mathrm{c}}$. More generally, a better understanding of other response
functions besides the conductivity, such as the magnetic susceptibility,
would be highly desirable. The response is expected to be strongly affected
by disorder. Understanding the spin and charge transport and the optical
response of DMS might be the most challenging but, in view of possible
applications, the most worthwhile goal.

\ack

Stimulating discussions with W. A. Atkinson, M. Berciu, T. Dietl, S. C.
Erwin, G. A. Fiete, F. H\"of\-ling, P. Kacman, J. K\"onig, L. W. Molenkamp,
M. E. Raikh, F. Sch\"afer, J. Schliemann, M. B. Silva Neto, J. Sinova, G.
Zar\'and and in particular with F. von Oppen and A. H. MacDonald are
gratefully acknowledged. The author also thanks the University of Texas at
Austin for hospitality and the Deutsche Forschungsgemeinschaft for
financial support.

\end{document}